\documentclass[preprint,prd,aps,showpacs,showkeys,nofootinbib]{revtex4}
\usepackage{graphicx}
\usepackage{dcolumn}
\usepackage{bm}
\usepackage{color}
\usepackage[dvipsnames]{xcolor}
\usepackage{amssymb}

\definecolor{light-gray}{gray}{0.88}
\definecolor{dark-gray}{gray}{0.40}
\begin{document}
\title{Study lepton flavor violation
$B^0\rightarrow{{l_i}^{\pm}{l_j}^{\mp}}$
 within the Mass Insertion Approximation}
\author{Yi-Tong Wang$^{1,2,3}$, Jiao Ma$^{1,2,3}$, Xing-Yu Han$^{1,2,3}$,  Xin-Xin Long$^{1,2,3}$, Tong-Tong Wang$^{1,2,3}$, Hai-Bin Zhang$^{1,2,3}$, Shu-Min Zhao$^{1,2,3}$\footnote{zhaosm@hbu.edu.cn}}

\affiliation{$^1$ Department of Physics, Hebei University, Baoding 071002, China}
\affiliation{$^2$ Hebei Key Laboratory of High-precision Computation and Application of Quantum Field Theory, Baoding, 071002, China}
\affiliation{$^3$ Hebei Research Center of the Basic Discpline for Computational Physics, Baoding, 071002, China}

\date{\today}

\begin{abstract}
We study lepton flavor violating (LFV) decays $B^0\rightarrow{{l_i}^{\pm}{l_j}^{\mp}}$ ($B^0\rightarrow e{\mu}$, $B^0\rightarrow e{\tau}$ and $B^0\rightarrow {\mu}{\tau}$) in the $U(1)_X$SSM (SSM is acronym for the supersymmetric standard model), which is the $U(1)_X$ extension of the minimal supersymmetric standard model (MSSM). The local gauge group of $U(1)_X$SSM is $SU(3)_C\times SU(2)_L \times U(1)_Y \times U(1)_X$. These processes ($B^0\rightarrow e{\mu}$, $B^0\rightarrow e{\tau}$ and $B^0\rightarrow {\mu}{\tau}$) are strictly forbidden in the standard model (SM), but these LFV decays are a signal of new physics (NP).
We use the Mass Insertion Approximation (MIA) to find sensitive parameters that directly influence the result of the branching ratio of LFV decay $B_d\rightarrow{{l_i}^{\pm}{l_j}^{\mp}}$. Combined with the latest experimental results, we analyze the relationship between different sensitive parameters and the branching ratios of the three processes. According to the numerical analysis, these elements are very restricted from the non observation of CLFV.

\end{abstract}


\maketitle

\section{Introduction}

Certain processes are sensitive to their potential contributions in the new physics (NP) model but are suppressed or prohibited in the standard model (SM). The lepton flavor violation (LFV) decays are forbidden in the SM, and neutrino oscillation experiments show that lepton flavor symmetry is not conserved in the neutrino sector\cite{IN0,IN1,INj2,INj3}. If LFV is detected in the charged sector,  its observation will be clear evidence of physics beyond SM. Due to various experiments to find LFV decay, the LFV decays have recently been discussed within various theoretical frameworks\cite{B11,B12,B13,B14,B15,B16,B17,B18,B19}.
SM has been surprisingly successful in interpreting data, but SM is still considered incomplete.
It still has several problems that cannot be solved, and in order to solve these open puzzles, it is essential to explore the physics beyond SM (BSM). In this case, B decay becomes a good testing ground for exploring new physics (NP) beyond SM through low-energy experiments. The search for LFV in B decays has also been ongoing\cite{tj1,tj2,tj3}. Another motivation for exploring the LFV B decays comes from an anomaly in the measurement of the lepton flavor universality ratio associated with $B\rightarrow s{\mu^+}{\mu^-}$ (see Ref.\cite{tj4}). Models have been independently proposed \cite{tj1,tj5,tj6}, and the interpretation of the data tends to predict large rates of decay of LFV B, especially in the $\mu-\tau$ sector.
The LFV decays $B^0\rightarrow{{l_i}^{\pm}{l_j}^{\mp}}$ ($B^0\rightarrow e{\mu}$, $B^0\rightarrow e{\tau}$ and $B^0\rightarrow {\mu}{\tau}$) are interesting, and are helpful for our exploration of NP beyond the SM.

The decays of LFV for B-meson are forbidden in the SM with massless neutrino. However, if the SM is minimally extended to explain the observed neutrino oscillations.
The expected branching ratio of process $B^0\rightarrow {\mu}{\tau}$ is of the order $10^{-54}$ as
shown in Ref.\cite{IN2}, which is well below current and expected experimental sensitivity. These
processes have been similarly explored in several other new physics models. In the leptoquarks\cite{T3,T8} or new specification $Z^{'}$ boson model\cite{T7}, the branching ratio of the $B^0\rightarrow e{\mu}$ process can reach $10^{-11}$. Possible enhancements to the process of $B^0\rightarrow e{\mu}$ is also predicted in other models, such as heavy singlet Dirac neutrinos\cite{T9}, and the Pati-Salam model\cite{T4}.

For the LFV decay $B^0\rightarrow e{\tau}$, some new physical models give the branching ratio of $10^{-9}$ to $10^{-10}$, such as the Pati-Salam vector leptoquarks with the mass of 86 ${\rm TeV}$ give the branching ratio of $1.6\times10^{-9}$\cite{T5}. For the LFV decay $B^0\rightarrow {\mu}{\tau}$, the branching ratio can reach $10^{-8}$ in the model containing the heavy neutral gauge $Z^{'}$ boson\cite{Z4}. In scalar or vector leptoquarks models, the prediction range of the branching ratio of $B^0\rightarrow {\mu}{\tau}$ is given as $10^{-9}-10^{-5}$\cite{T8,T5,Z1}. These decay processes have previously been researched by the CLEO\cite{B21}, Belle\cite{B22} and LHCb\cite{B23,B24} experiments. But no evidence has been observed so far. The current strictest experimental limits are given in the Table \ref{I}\cite{B25,B26,B27}. The latest upper limits on the branching ratios of $B^0\rightarrow{M^0{l_i}^{\pm}{l_j}^{\mp}}$ and $B^+\rightarrow{M^+{l_i}^{\pm}{l_j}^{\mp}}$, with $M = \pi, \rho$ at 90\% confidence level (CL.) are given in the Table \ref{I1}\cite{B25,ff1,ff3,ff4}. In the study of the Drell-Yan process using LHC data at high-$p_T$ conditions, the upper limits of the branching ratios of process $B^0\rightarrow {\rho^0}e{\tau}$ and process $B^0\rightarrow {\rho^0}{\mu}{\tau}$ can reach $1.4\times10^{-4}$ and $7.5\times10^{-5}$ at 95\% confidence level (CL.), and the expected sensitivity on the HL-LHC can be $3.0\times10^{-5}$ and $1.5\times10^{-5}$\cite{ff5}.

In order to search for LFV in charged leptons, many experiments are currently planned, under construction, or already underway, with the aim of increasing the current limits by several orders of magnitude. Some of the most stringent constraints for LFV processes are given in Refs.\cite{tj7,tj8,tj9,tj10,tj11,tj12}.
In this article, the B-decays  represent the decay processes $B^0\rightarrow{{l_i}^{\pm}{l_j}^{\mp}}$ ($B^0\rightarrow e{\mu}$, $B^0\rightarrow e{\tau}$ and $B^0\rightarrow {\mu}{\tau}$).
The searches for charged lepton flavor violating decays should be categorised into modes with ${\tau}$ leptons, and modes without, i.e. with an electron and a muon. Limits at the few $10^{-9}$ level exist for process $B^0\rightarrow e{\mu}$\cite{B23}, and will be improved by Belle II and LHCb. An improvement by an order of magnitude by the end of LHCb Upgrade II is reachable. Similar LFV searches have
been performed with charm\cite{ff4,B28,FB1}.

The LFV decay of B-meson provides a finer probe for finding NP.
In addition to the process studied in this paper, the B-meson decay process $B_s\rightarrow{{l_i}^{\pm}{l_j}^{\mp}}$ is also very interesting.
Over the past decade, we have witnessed tremendous efforts by the high-energy physics community to better understand exclusive decays based on $B\rightarrow sll$, as their detailed experimental analysis became feasible at the Large Hadron Collider (LHC)\cite{tj13,tj14,tj15,tj16,tj17}. $B\rightarrow sll$ decay is thought to be the main window to the particle content of physical BSM.
The symbol Br represents the branching ratio.
The LHCb experiment sets the latest upper limits on $Br(B_s\rightarrow e{\mu})$ as $5.4\times10^{-9}$, and $Br(B_s\rightarrow {\mu}{\tau})$ as
$3.4\times10^{-5}$ at 90\% confidence level (CL.)\cite{B23,B24}.
In the exploration of some other new physical models, it can be observed that the branching ratio of the process $B_s\rightarrow e{\mu}$ appears in $10^{-9}$,  which is about four orders smaller than  $Br(B_s\rightarrow e{\tau})$ and $Br(B_s\rightarrow{\mu}{\tau})$\cite{tj3}. The NP contributions can be associated with new particles with mass scales well above the energy range of the LHC, for example, by a multi-TeV-scale $Z^{'}$ boson or a leptoquark. Therefore, the precise determination of the effective coupling through the measurement processes $B^0\rightarrow{{l_i}^{\pm}{l_j}^{\mp}}$ ($B^0\rightarrow e{\mu}$, $B^0\rightarrow e{\tau}$ and $B^0\rightarrow {\mu}{\tau}$) are essential to understand or constrain the structure of any NP model. These decay channels can be further analyzed at the upcoming LHC and B-factories, they may be events that could lead to the origin of single signals in NP.

$U(1)_X$SSM is the $U(1)_X$ extension of the MSSM and its local gauge group is $SU(3)_C\times SU(2)_L \times U(1)_Y \times U(1)_X$\cite{Sarah1,Sarah2,Sarah3}. On the basis of MSSM we add three new singlet Higgs superfields $\hat{\eta},~\hat{\bar{\eta}},~\hat{S}$ and three generations of right-handed neutrinos $\hat{\nu}_i$ to get $U(1)_X$SSM. The right-handed neutrinos have two functions, one of which is to produce  tiny mass to light neutrinos through see-saw mechanism. Another is to provide a new dark matter candidate-light sneutrino.
This model alleviates the so-called little hierarchy problem that occurs in MSSM. Compared with the MSSM, the neutrino masses in the $U(1)_X$SSM are not zero. The lightest CP-even Higgs mass at tree level is improved.
In $U(1)_X$SSM, the interaction between three extra singlet Higgs superfields and two Higgs doublets is favorable to increase the
mass of the lightest CP-even Higgs at the tree level.
The second light neutral CP-even Higgs can be at TeV order. In this article, we  analyze the LFV decays $B^0\rightarrow{{l_i}^{\pm}{l_j}^{\mp}}$ ($B^0\rightarrow e{\mu}$, $B^0\rightarrow e{\tau}$ and $B^0\rightarrow {\mu}{\tau}$) with the method of Mass Insertion Approximation (MIA)\cite{MIA1,T10,MIA2,MIA3,MIA4} in the $U(1)_X$SSM. The MIA is the calculation method that can more intuitively and clearly find out the sensitive parameters for LFV decays. It uses the characteristic states of electroweak interaction, and the perturbed mass insertion changes slepton flavor. What is noteworthy is that these parameters are considered between all possible flavor blends among SUSY partner of leptons, in which their particular origin has no assumption and is independent of the model\cite{MIA2}. The MIA method has also been applied to other work related to LFV\cite{MIA2,MIA3,MIA4}, which lays a good foundation for us to continue exploring LFV.
\begin{table}
\caption{ Current upper limits on LFV B decays considered in our analysis.}
\begin{tabular}{c|c|c}
\hline
decay modes &  \hspace{0.5cm}Upper Limit(90\%CL.)\hspace{0.5cm}  & \hspace{0.8cm} Future sensitivity\hspace{0.8cm} \\
\hline
$B^0\rightarrow e{\mu}$ & $1.0\times10^{-9}$\cite{B25,B26} & $9.0\times10^{-11}$\cite{B27} \\
\hline
$B^0\rightarrow e{\tau}$ & $1.6\times10^{-5}$\cite{B25,B26} & $-$ \\
\hline
$B^0\rightarrow {\mu}{\tau}$ & $1.2\times10^{-5}$\cite{B26} & $3.0\times10^{-6}$\cite{B27} \\
\hline
\end{tabular}
\label{I}
\end{table}

\begin{table}
\caption{ The current experimental limits on the relevant processes at 90\% confidence level (CL.). The dash symbol represent decays that are
kinematically allowed, but for which there is not an experimental limit yet. }
\begin{tabular}{c|c||c|c||c|c}
\hline
decay modes &  \hspace{0.5cm}Upper Limit\hspace{0.5cm}  & decay modes &  \hspace{0.5cm}Upper Limit\hspace{0.5cm} & decay modes &  \hspace{0.5cm}Upper Limit\hspace{0.5cm} \\
\hline
$B^+\rightarrow {\pi^+}e{\mu}$ & $1.7\times10^{-7}$\cite{B25,ff1} & $B^+\rightarrow {\pi^+}e{\tau}$ & $7.5\times10^{-5}$\cite{ff3,ff4} & $B^+\rightarrow {\pi^+}{\mu}{\tau}$ & $7.2\times10^{-5}$\cite{ff3,ff4} \\
$B^0\rightarrow {\pi^0}e{\mu}$ & $1.4\times10^{-7}$\cite{B25,ff1} & $B^0\rightarrow {\pi^0}e{\tau}$ & - & $B^0\rightarrow {\pi^0}{\mu}{\tau}$ & - \\
$B^0\rightarrow {\rho^0}e{\mu}$ & - & $B^0\rightarrow {\rho^0}e{\tau}$ & - & $B^0\rightarrow {\rho^0}{\mu}{\tau}$ & - \\
\hline
\end{tabular}
\label{I1}
\end{table}

The direct search for particle production in collider experiments plays a crucial role in the search for supersymmetry (SUSY). In recent years, the results of direct searches for SUSY particles at the collider have mainly included data analysis from the ATLAS and CMS experiments, with reference to results from LEP, HERA and the Tevatron. In the parameters chosen in this paper, we use slepton as an example to directly search for the production of slepton in the collider experiment. In the LHC, the pair production of sleptons is not only severely suppressed compared to the pair production of colored SUSY particles, but the cross sections are also nearly two orders of magnitude smaller than those produced by pairs with charginos and neutralinos. ATLAS and CMS have searched for direct production of selectron pairs and smuon pairs at the LHC. In simplified models, ATLAS and CMS set lower mass limits
on sleptons of 700 GeV for degenerate $\tilde{l}_L$ and $\tilde{l}_R$\cite{tj3}.

In our previous works\cite{T10,T1}, we have researched the LFV decays $l_j\rightarrow{l_i\gamma}$ with the method of MIA and Mass Eigenstate in the $U(1)_X$SSM.
From our investigation, it can be found that the present experimental limit of $l_j\rightarrow{l_i\gamma}$ branching ratio strictly constrains the parameter space of $U(1)_X$SSM to a great extent. In this work, we still consider the constraints of the existing experimental
limits of the $l_j\rightarrow{l_i\gamma}$ branching ratios. The latest upper limits on the branching ratios of $l_j\rightarrow{l_i\gamma}$ (${\mu}\rightarrow{e\gamma}$, ${\tau}\rightarrow{e\gamma}$ and ${\tau}\rightarrow{{\mu}\gamma}$)
at 90\% confidence level (CL.)\cite{B25,tj11,tj19} are
\begin{eqnarray}
&&Br({\mu}\rightarrow{e\gamma})<4.2\times10^{-13}\!,~~Br({\tau}\rightarrow{e\gamma})<3.3\times10^{-8}\!,~~Br({\tau}\rightarrow{{\mu}\gamma})<4.4\times10^{-8}\!.
\end{eqnarray}
Under the flavor constraint of process $l_j\rightarrow{l_i\gamma}$, the decay of LFV $B^0$ will be much lower than the experimental sensitivity.

We work mainly on the following aspects. In Section II, we briefly introduce the main content of $U(1)_X$SSM, including its superpotential,
the general soft SUSY-breaking terms and couplings. In Section III, we give an analytical expression for the branching ratio of $B^0\rightarrow{{l_i}^{\pm}{l_j}^{\mp}}$ decay, and use MIA to calculate the analytical results and the degenerate results of $B^0\rightarrow{{l_i}^{\pm}{l_j}^{\mp}}$ in $U(1)_X$SSM. In Section IV, we perform numerical analysis for different parameter spaces, and give corresponding two-dimensional graphs and find the sensitive parameters. The discussion and conclusion are given in
Section V.
\section{The main content of $U(1)_X$SSM}
$U(1)_X$SSM is the $U(1)_X$ extension of MSSM, and the local gauge group is $SU(3)_C\otimes
SU(2)_L \otimes U(1)_Y\otimes U(1)_X$\cite{T1,UU1,UU3,UU4,SY1,UU5,UU6}.
Compared to MSSM, $U(1)_X$SSM has new superfields, such as right-handed neutrinos $\hat{\nu}_i$ and three Higgs singlets $\hat{\eta},~\hat{\bar{\eta}},~\hat{S}$.
Through the seesaw mechanism, light neutrinos acquire tiny mass at the tree level.
The neutral CP-even parts of $H_u,~ H_d,~\eta,~\bar{\eta}$ and $S$ mix together, forming $5\times5 $ mass squared matrix.
The loop corrections to the lightest CP-even Higgs mass are important, because we need it to get the Higgs mass of 125.25 GeV\cite{B25}. The sneutrinos are disparted into CP-even sneutrinos and CP-odd sneutrinos, and their mass squared matrixes are both extended to $6\times6$.

 In $U(1)_X$SSM, the superpotential is expressed as:
\begin{eqnarray}
&&W=l_W\hat{S}+\mu\hat{H}_u\hat{H}_d+M_S\hat{S}\hat{S}-Y_d\hat{d}\hat{q}\hat{H}_d-Y_e\hat{e}\hat{l}\hat{H}_d+\lambda_H\hat{S}\hat{H}_u\hat{H}_d
\nonumber\\&&~~~~~~+\lambda_C\hat{S}\hat{\eta}\hat{\bar{\eta}}+\frac{\kappa}{3}\hat{S}\hat{S}\hat{S}+Y_u\hat{u}\hat{q}\hat{H}_u+Y_X\hat{\nu}\hat{\bar{\eta}}\hat{\nu}
+Y_\nu\hat{\nu}\hat{l}\hat{H}_u.
\end{eqnarray}

 The specific explicit expressions of two Higgs doublets are as follows:
\begin{eqnarray}
&&H_{u}=\left(\begin{array}{c}H_{u}^+\\{1\over\sqrt{2}}\Big(v_{u}+H_{u}^0+iP_{u}^0\Big)\end{array}\right),
~~~~~~
H_{d}=\left(\begin{array}{c}{1\over\sqrt{2}}\Big(v_{d}+H_{d}^0+iP_{d}^0\Big)\\H_{d}^-\end{array}\right).
\end{eqnarray}

 The three Higgs singlets are represented by:
\begin{eqnarray}
&&\eta={1\over\sqrt{2}}\Big(v_{\eta}+\phi_{\eta}^0+iP_{\eta}^0\Big),~~~~~~~~~~~~~~~
\bar{\eta}={1\over\sqrt{2}}\Big(v_{\bar{\eta}}+\phi_{\bar{\eta}}^0+iP_{\bar{\eta}}^0\Big),\nonumber\\&&
\hspace{3.0cm}S={1\over\sqrt{2}}\Big(v_{S}+\phi_{S}^0+iP_{S}^0\Big).
\end{eqnarray}

Here,  $v_u,~v_d,~v_\eta$,~ $v_{\bar\eta}$ and $v_S$ are the corresponding vacuum expectation values(VEVs) of the Higgs superfields $H_u$, $H_d$, $\eta$, $\bar{\eta}$ and $S$.
Two angles are defined as $\tan\beta=v_u/v_d$ and $\tan\beta_\eta=v_{\bar{\eta}}/v_{\eta}$.
The definition of ${\widetilde{\nu}}_{L}$ and ${\widetilde{\nu}}_{R}$ are:
\begin{eqnarray}
&&\widetilde{\nu}_{L}= \frac{1}{\sqrt{2} } {\phi}_{l}+\frac{i}{\sqrt{2} } {\sigma}_{l},~~~~~~~~~~~~~~
\widetilde{\nu}_{R}= \frac{1}{\sqrt{2} } {\phi}_{R}+\frac{i}{\sqrt{2} } {\sigma}_{R}.
\end{eqnarray}

 The soft SUSY breaking terms of $U(1)_X$SSM are:
\begin{eqnarray}
&&\mathcal{L}_{soft}=\mathcal{L}_{soft}^{MSSM}-B_SS^2-L_SS-\frac{T_\kappa}{3}S^3-T_{\lambda_C}S\eta\bar{\eta}
+\epsilon_{ij}T_{\lambda_H}SH_d^iH_u^j\nonumber\\&&\hspace{1.5cm}
-T_X^{IJ}\bar{\eta}\tilde{\nu}_R^{*I}\tilde{\nu}_R^{*J}
+\epsilon_{ij}T^{IJ}_{\nu}H_u^i\tilde{\nu}_R^{I*}\tilde{l}_j^J
-m_{\eta}^2|\eta|^2-m_{\bar{\eta}}^2|\bar{\eta}|^2-m_S^2S^2\nonumber\\&&\hspace{1.5cm}
-(m_{\tilde{\nu}_R}^2)^{IJ}\tilde{\nu}_R^{I*}\tilde{\nu}_R^{J}
-\frac{1}{2}\Big(M_X\lambda^2_{\tilde{X}}+2M_{BB^\prime}\lambda_{\tilde{B}}\lambda_{\tilde{X}}\Big)+h.c~~.
\end{eqnarray}

 The particle content and charge distribution of $U(1)_X$SSM are shown in the Table \ref {IV}. In previous work\cite{UU4}, we have shown that $U(1)_X$SSM is anomaly free. The details regarding the absence of anomaly within the $U(1)_X$SSM model can be summarized as follows:

1. Similar as the SM condition, the anomaly of three $SU(2)_L$($SU(3)_C$) gauge bosons vanishes.

2. The anomaly caused by one $SU(3)_C$ boson or one $SU(2)_L$ boson is proportional to $Tr[t^a]=0$ or $Tr[\tau ^a]=0$.

3. The anomaly of one $U(1)_Y$ boson with two $SU(3)_C$ bosons is proportional
to the factor $Tr[t^at^bY^Y ] =\frac{1}{2}{\delta ^{ab}} \Sigma_q Y^Y_q$.
The anomaly of one $U(1)_X$ boson with two $SU(3)_C$ bosons is proportional
to the  factor $Tr[t^at^bY^X] = \frac{1}{2}{\delta ^{ab}} \Sigma_q Y^Y_q$.

4. $Tr[\tau^a\tau^bY^Y] = \frac{1}{2}{\delta ^{ab}} \Sigma_L Y^Y_L$ corresponds to the anomaly of one $U(1)_Y$ boson with two $SU(2)_L$ bosons.
While, $Tr[\tau^a\tau^bY^X] = \frac{1}{2}{\delta ^{ab}} \Sigma_L Y^X_L$ corresponds to the anomaly of one $U(1)_X$ boson with two $SU(2)_L$ bosons.

5. The anomalies of three U(1) gauge bosons are shown in the following form
\begin{eqnarray}
&&Tr[Y^YY^YY^Y ]= \sum_{n}(Y^Y_n)^3, ~~~~~Tr[Y^XY^XY^X] =\sum_{n}(Y^X_n)^3,\nonumber\\
&&Tr[Y^XY^YY^Y ]= \sum_{n}Y^X_n(Y^Y_n)^2, ~~~~T r[Y^YY^XY^X] =\sum_{n}Y^Y_n(Y^X_n)^2.
\end{eqnarray}

6. The gravitational anomaly with one $U(1)_Y$ gauge boson is proportional to $Tr[Y^Y ] =\Sigma_n Y^Y_n$.
The gravitational anomaly with one $U(1)_X$ gauge boson is proportional to $Tr[Y^X ] =\Sigma_n Y^X_n$.

The two Abelian groups $U(1)_Y$ and $U(1)_X$ in $U(1)_X$SSM cause a new effect: the gauge kinetic mixing.
This effect can be induced by the renormalization group equations (RGEs).

\begin{table}
\caption{ The superfields in $U(1)_X$SSM}
\begin{tabular}{|c|c|c|c|c|c|c|c|c|c|c|c|}
\hline
Superfields & $\hspace{0.1cm}\hat{q}_i\hspace{0.1cm}$ & $\hat{u}^c_i$ & $\hspace{0.2cm}\hat{d}^c_i\hspace{0.2cm}$ & $\hat{l}_i$ & $\hspace{0.2cm}\hat{e}^c_i\hspace{0.2cm}$ & $\hat{\nu}_i$ & $\hspace{0.1cm}\hat{H}_u\hspace{0.1cm}$ & $\hat{H}_d$ & $\hspace{0.2cm}\hat{\eta}\hspace{0.2cm}$ & $\hspace{0.2cm}\hat{\bar{\eta}}\hspace{0.2cm}$ & $\hspace{0.2cm}\hat{S}\hspace{0.2cm}$ \\
\hline
$SU(3)_C$ & 3 & $\bar{3}$ & $\bar{3}$ & 1 & 1 & 1 & 1 & 1 & 1 & 1 & 1  \\
\hline
$SU(2)_L$ & 2 & 1 & 1 & 2 & 1 & 1 & 2 & 2 & 1 & 1 & 1  \\
\hline
$U(1)_Y$ & 1/6 & -2/3 & 1/3 & -1/2 & 1 & 0 & 1/2 & -1/2 & 0 & 0 & 0  \\
\hline
$U(1)_X$ & 0 & -1/2 & 1/2 & 0 & 1/2 & -1/2 & 1/2 & -1/2 & -1 & 1 & 0  \\
\hline
\end{tabular}
\label{IV}
\end{table}

The general form of the covariant derivative of $U(1)_X$SSM can be found in Refs.\cite{UMSSM5,B-L1,B-L2,gaugemass}. In $U(1)_X$SSM, the gauge bosons $A^{X}_\mu,~A^Y_\mu$ and $V^3_\mu$ mix together at the tree level. The mass matrix
in the basis $(A^Y_\mu, V^3_\mu, A^{X}_\mu)$ can be found in Ref.\cite{UU4}. We use two mixing angles $\theta_{W}$ and $\theta_{W}'$ to obtain mass eigenvalues of the matrix. $\theta_{W}$ is the Weinberg angle and $\theta_{W}'$ is the new mixing angle. The new mixing angle is defined as
\begin{eqnarray}
\sin^2\theta_{W}'\!=\!\frac{1}{2}\!-\!\frac{[(g_{{YX}}+g_{X})^2-g_{1}^2-g_{2}^2]v^2+
4g_{X}^2\xi^2}{2\sqrt{[(g_{{YX}}+g_{X})^2+g_{1}^2+g_{2}^2]^2v^4\!+\!8g_{X}^2[(g_{{YX}}+g_{X})^2\!-\!g_{1}^2\!-\!g_{2}^2]v^2\xi^2\!+\!16g_{X}^4\xi^4}}.
\end{eqnarray}
Here, $v^2=v_u^2+v_d^2$ and $\xi^2=v_\eta^2+v_{\bar{\eta}}^2$.

 The new mixing angle appears in the couplings involving $Z$ and $Z^{\prime}$.
The exact eigenvalues are calculated
\begin{eqnarray}
&&m_\gamma^2=0,\nonumber\\
&&m_{Z,{Z^{'}}}^2=\frac{1}{8}\Big([g_{1}^2+g_2^2+(g_{{YX}}+g_{X})^2]v^2+4g_{X}^2\xi^2 \nonumber\\
&&\hspace{1.1cm}\mp\sqrt{[g_{1}^2+g_{2}^2+(g_{{YX}}+g_{X})^2]^2v^4\!+\!8[(g_{{YX}}+g_{X})^2\!-\!g_{1}^2\!-\!
g_{2}^2]g_{X}^2v^2\xi^2\!+\!16g_{X}^4\xi^4}\Big).
\end{eqnarray}
The used mass matrices can be found in the works\cite{UU1,T1}.

Here, we show some couplings that we need in the $U(1)_X$SSM:
\begin{eqnarray}
&&\mathcal{L}_{\bar{l}\chi^-\tilde{\nu}^R}=\frac{1}{\sqrt{2}}\bar{l}_i\Big\{\tilde{\nu}^R_{L,i}Y_l^iP_L\tilde{H}^-_d
-g_2\tilde{\nu}^R_{L,i}P_R\tilde{W}^-\Big\},\nonumber\\
&&\mathcal{L}_{\bar{l}\chi^-\tilde{\nu}^I}=\frac{1}{\sqrt{2}}\bar{l}_i\Big\{\tilde{\nu}^I_{L,i}Y_l^iP_L\tilde{H}^-_d
-g_2\tilde{\nu}^I_{L,i}P_R\tilde{W}^-\Big\},
%%%%%%%%%%%%%%%%%%%%%%%%%%%%%%%%%%%%%%%%%
\\&&\mathcal{L}_{\bar{\chi}^0l\tilde{e}}=\Big\{\Big(\frac{1}{\sqrt{2}}(g_1 \tilde{B}+g_2 \tilde{W}^0+g_{YX}\lambda_{\tilde{X}})\tilde{e}_j^L-\tilde{H}_d^0Y_l^j\tilde{e}_j^R\Big)P_L\nonumber\\
&&\hspace{1.5cm}-\Big[\frac{1}{\sqrt{2}}(2g_1\tilde{B}+(2g_{YX}+g_X)\lambda_{\tilde{X}})\tilde{e}_j^R+\tilde{H}_d^0Y_l^j\tilde{e}_j^L\Big]P_R\Big\}\l_j,
%%%%%%%%%%%%%%%%%%%%%%%%%%%%%%%%%%%%%%%%%
\\&&\mathcal{L}_{\bar{\chi}^-d\tilde{U}}=\Big[\Big(\tilde{H}^+_uY_{u,i}\tilde{U}^{R*}_i-g_2\tilde{W}^-\tilde{U}^{L*}_i)P_L+Y_{d,j}\tilde{U}^{L*}_i\tilde{H}^-_dP_R\Big]V_{ij}d_j,
%%%%%%%%%%%%%%%%%%%%%%%%%%%%%%%%%%%%%%%%%
\\&&
\mathcal{L}_{\bar{d}\chi^0\tilde{D}}=\bar{d}_{i}
\Big\{-\frac{1}{6} \Big[\sqrt{2}\tilde{D}_{i}^R(2 g_1 \tilde{B} + (2g_{YX}+3g_X)\lambda_{\tilde{X}}) + 6Y_{d,i} \tilde{H}_d^0\tilde{D}_i^L\Big]P_L
\nonumber\\&&\hspace{1.5cm}- \frac{1}{6} \Big[6 Y_{d,i}\tilde{D}_i^R\tilde{H}_d^0 + \sqrt{2}\tilde{D}_i^L(g_1 \tilde{B} + g_{YX} \lambda_{\tilde{X}} - 3 g_2\tilde{W}^0 ) \Big]P_R \Big\}.
\end{eqnarray}

\section{ FORMULATION }
 In this section, we focus on the LFV of the $B^0\rightarrow{{l_i}^{\pm}{l_j}^{\mp}}$ with the MIA under the $U(1)_X$SSM model. The most structured effective Hamiltonian describing the $B^0\rightarrow{{l_i}^{\pm}{l_j}^{\mp}}$ process can be expressed as
\begin{eqnarray}
\mathcal {H}_{eff}(B^0\rightarrow{{l_i}^{\pm}{l_j}^{\mp}})=-\frac{4G_F}{\sqrt{2}}V_{td}^*V_{tb}
\sum_{i=1}^5C_i\mathcal{O}_i,
\label{eff-hamiltonian}\label{G8}
\end{eqnarray}
where $V_{td}^*V_{tb}$ is the CKM matrix element. For Eq.(\ref{G8}), $C_i$ stands for Wilson coefficients. The effective operators relevant to our study are:
\begin{eqnarray}
&&\mathcal{O}_1=(\overline{l}_i\gamma^{\mu}P_Ll_j)(\overline{b}\gamma_{\mu}P_Ld),~~~~\mathcal{O}_2=(\overline{\tau}P_Ll_i)(\overline{b}P_Rd),\nonumber\\
&&\mathcal{O}_3=(\overline{\tau}\gamma^{\mu}P_Ll_i)(\overline{b}\gamma_{\mu}P_Ld),~~~~~\mathcal{O}_4=(\overline{l}_iP_R\tau)(\overline{b}P_Ld),\nonumber\\
&&\hspace{1.9cm}\mathcal{O}_5=(\overline{l}_i\gamma^{\mu}P_L\tau)(\overline{b}\gamma_{\mu}P_Ld).
\end{eqnarray}
The branching ratios of $B^0\rightarrow{{l_i}^{\pm}{l_j}^{\mp}}$ are defined as
\begin{eqnarray}
&&BR(B^0\rightarrow{{l_i}^{\pm}{l_j}^{\mp}})=\tau_{B^0} \frac{G_F^2}{16\pi}
\frac{f_{B^0}^2}{m_{B^0}}\lambda^{\frac{1}{2}}(m_{B^0}^2,m_{l_i}^2,m_{l_j}^2)\nonumber\\
&&\hspace{3.5cm}\times\Big[(m_{B^0}^2-(m_{l_i}+m_{l_j})^2)|(m_{l_i}-m_{l_j})A+\frac{m_{B^0}^2}{m_b+m_d}C|^2\nonumber\\
&&\hspace{3.5cm}+(m_{B^0}^2-(m_{l_i}-m_{l_j})^2)|(m_{l_i}+m_{l_j})B+\frac{m_{B^0}^2}{m_b+m_d}C|^2\Big],
\end{eqnarray}
where $\lambda(\alpha,\beta,\gamma)=\alpha^2+\beta^2+\gamma^2-2\alpha\beta-2\alpha\gamma-2\beta\gamma$
and $f_{B^0}$ is the $B^0$-meson decay constant. For the process of $B^0\rightarrow {e\mu}$, $A=-C_1$,
$B=C_1$, $C=0$; for the processes of $B^0\rightarrow {e\tau}$ and $B^0\rightarrow {\mu\tau}$,
$A=-(C_1+C_3+C_5)$, $B=C_1+C_3+C_5$, $C=-C_2-C_4$.

\subsection{Using MIA to calculate $B^0\rightarrow{{l_i}^{\pm}{l_j}^{\mp}}$ in $U(1)_X$SSM}
\begin{figure}[h]
\setlength{\unitlength}{5.0mm}
\centering
\includegraphics[width=5.0in]{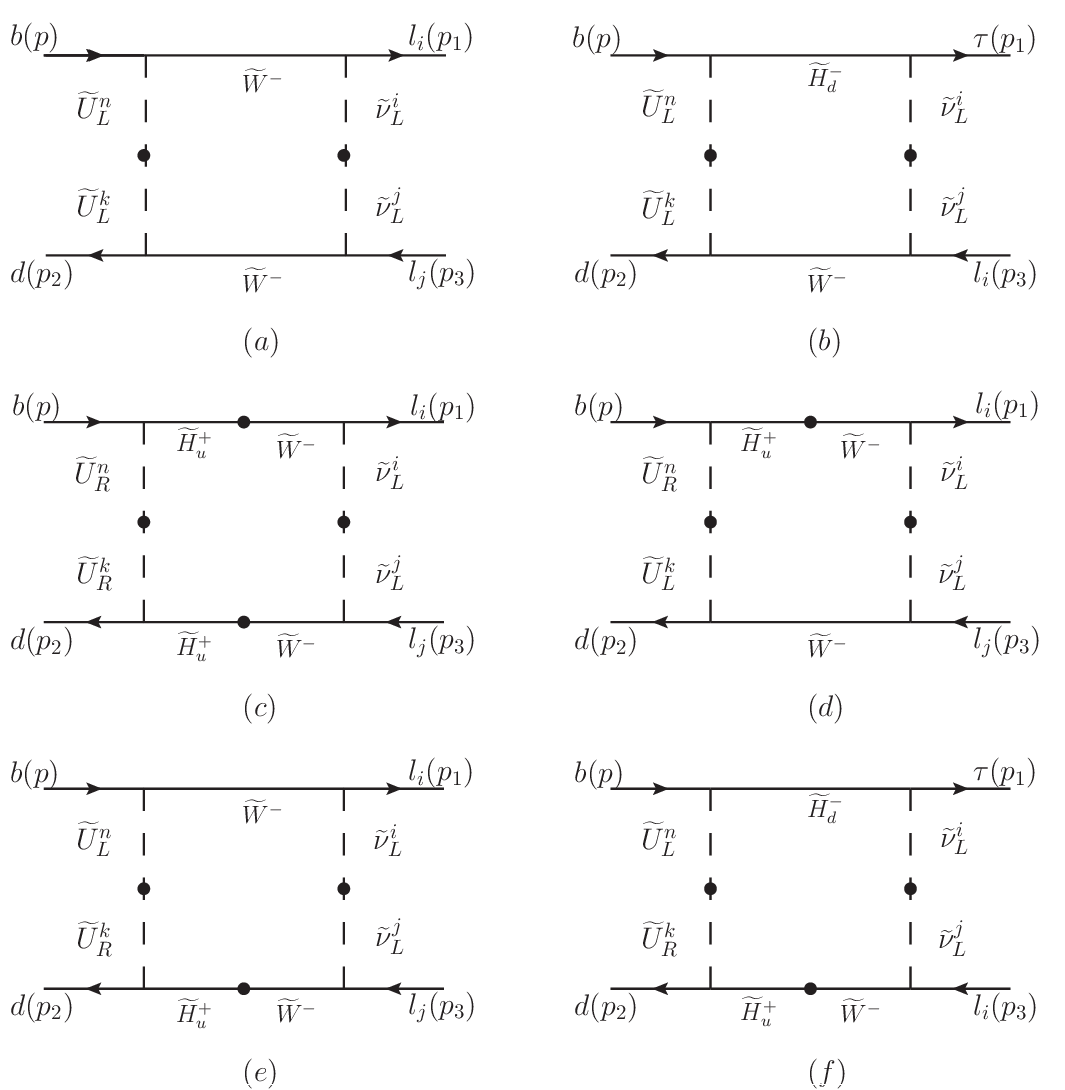}
\caption{The box-type diagrams for the $B^0\rightarrow{{l_i}^{\pm}{l_j}^{\mp}}$ processes of $\chi^c$ in the MIA.}\label{N1}
\end{figure}

The box-type diagrams drawn in Fig.1 can be written as
\begin{eqnarray}
&&T_{box}^1=(B_1^1+B_1^3+B_1^4+B_1^5)\overline{l}_i(p_1)\gamma^{\mu}P_Ll_j(p_3)\overline{b}(p)\gamma_{\mu}P_Ld(p_2)\nonumber\\
&&\hspace{1.2cm}+(B_1^2+B_1^6)\overline{\tau}(p_1)P_Ll_i(p_3)\overline{b}(p)P_Rd(p_2).
\end{eqnarray}

From the box-type diagrams, we obtain the contribution of charged particles to effective couplings $B_1^\alpha$ with $\alpha=1...6$
\begin{eqnarray}
&&B_1^1=\sum_{n,k=1}^{3}\frac{1}{8\mathcal{H}\Lambda^6}\Delta^{LL}_{nk}(\widetilde{U})\Delta^{LL}_{ij}(\widetilde{\nu})g_2^4V_{3n}V_{1k}^{*}
I_2(x_2,x_{\widetilde{U}^n_L},x_{\widetilde{U}^k_L},x_{\widetilde{\nu}^i_L},x_{\widetilde{\nu}^j_L}),\nonumber\\
&&B_1^2=\sum_{n,k=1}^{3}\frac{1}{2\mathcal{H}\Lambda^6}\frac{g_2^2m_bm_{\tau}}{v^2\cos^2\beta}\Delta^{LL}_{nk}(\widetilde{U})\Delta^{LL}_{ij}(\widetilde{\nu})V_{3n}V_{1k}^{*}
I_3(x_{\mu_H^{'}},x_2,x_{\widetilde{U}^n_L},x_{\widetilde{U}^k_L},x_{\widetilde{\nu}^i_L},x_{\widetilde{\nu}^j_L}),\nonumber\\
&&B_1^3=\sum_{n,k=1}^{3}\frac{1}{2\mathcal{H}\Lambda^8}\frac{g_2^2m_W^2m_{u^n}m_{u^k}}{v^2\sin^2\beta}\Delta^{RR}_{nk}(\widetilde{U})\Delta^{LL}_{ij}(\widetilde{\nu})V_{3n}V_{1k}^{*}
\big(\sqrt{x_2x_{\mu_H^{'}}}+\sin\beta \cos\beta(x_2\nonumber\\
&&\hspace{1.0cm}+x_{\mu_H^{'}})\big) I_5(x_{\mu_H^{'}},x_2,x_{\widetilde{U}^n_R},x_{\widetilde{U}^k_R},x_{\widetilde{\nu}^i_L},x_{\widetilde{\nu}^j_L}),\nonumber\\
&&B_1^4=-\sum_{n,k=1}^{3}\frac{1}{4\mathcal{H}\Lambda^7}\frac{g_2^3m_Wm_{u^n}}{v\sin\beta}\Delta^{RL}_{nk}(\widetilde{U})\Delta^{LL}_{ij}(\widetilde{\nu})V_{3n}V_{1k}^{*}
(\sqrt{x_{\mu_H^{'}}}\cos\beta+\sqrt{x_2}\sin\beta)\nonumber\\
&&\hspace{1.0cm}\times I_4(x_2,x_{\mu_H^{'}},x_{\widetilde{U}^n_R},x_{\widetilde{U}^k_L},x_{\widetilde{\nu}^i_L},x_{\widetilde{\nu}^j_L}),\nonumber\\
&&B_1^5=-\sum_{n,k=1}^{3}\frac{1}{4\mathcal{H}\Lambda^7}\frac{g_2^3m_Wm_{u^k}}{v\sin\beta}\Delta^{LR}_{nk}(\widetilde{U})\Delta^{LL}_{ij}(\widetilde{\nu})V_{3n}V_{1k}^{*}
(\sqrt{x_2}\cos\beta+\sqrt{x_{\mu_H^{'}}}\sin\beta)\nonumber\\
&&\hspace{1.0cm}\times I_4(x_2,x_{\mu_H^{'}},x_{\widetilde{U}^n_L},x_{\widetilde{U}^k_R},x_{\widetilde{\nu}^i_L},x_{\widetilde{\nu}^j_L}),\nonumber\\
&&B_1^6=-\sum_{n,k=1}^{3}\frac{1}{\mathcal{H}\Lambda^7}\frac{g_2m_Wm_bm_{\tau}m_{u^k}}{v^3\cos^2\beta \sin\beta}\Delta^{LR}_{nk}(\widetilde{U})\Delta^{LL}_{ij}(\widetilde{\nu})
V_{3n}V_{1k}^{*}(\sqrt{x_2}\cos\beta+\sqrt{x_{\mu_H^{'}}}\sin\beta)\nonumber\\
&&\hspace{1.0cm}\times I_4(x_{\mu_H^{'}},x_2,x_{\widetilde{U}^n_L},x_{\widetilde{U}^k_R},x_{\widetilde{\nu}^i_L},x_{\widetilde{\nu}^j_L}).\nonumber\\
\end{eqnarray}
with
\begin{eqnarray}
&&I_1(x,y,z,w,t)=\frac{\rm i}{16\pi^2}\Big(\frac{x^2\log(x)}{(t\!-\!x)\! (x\!-\!w)\! (x\!-\!y)\! (x\!-\!z)}+\frac{y^2 \log (y)}{(t\!-\!y)\!(y\!-\!w)\!(y\!-\!x)\!(y\!-\!z)}\nonumber\\
&&\!+\frac{z^2 \log (z)}{(t\!-\!z)\!(z\!-\!w)\!(z\!-\!x)\!(z\!-\!y)}\!+\!\frac{w^2 \log (w)}{(t\!-\!w)\! (w\!-\!x) \!(w\!-\!y) \!(w\!-\!z)}\!-\!\frac{t^2 \log (t)}{(t\!-\!w)\!(t\!-\!x)\! (t\!-\!y)\! (t\!-\!z)}\Big),\nonumber\\
&&\hspace{4.0cm}I_2(x,y,z,w,t)=\frac{\partial{I_1(x,y,z,w,t)}}{\partial x},
%%%%%%%%%%%%%%%%%%%%%%%%%%%%%%%%%%%%%%%%%
\\&&\hspace{1.8cm}I_3(x,y,z,w,t,n)=\frac{\rm i}{16\pi^2}\Big(\frac{x^2\log(x)}{(n-x)\!(x-t)\!(x-w)\!(x-y)\!(x-z)}\nonumber\\
&&\hspace{0.5cm}+\frac{y^2\log(y)}{(n-y)\!(y-t)\!(y-w)\!(y-x)\!(y-z)}+\frac{z^2\log(z)}{(n-z)\!(z-t)\!(z-w)\!(z-x)\!(z-y)}\nonumber\\
&&\hspace{0.5cm}+\frac{w^2\log(w)}{(n-w)\!(w-t)\!(w-x)\!(w-y)\!(w-z)}+\frac{t^2\log(t)}{(n-t)\!(t-w)\!(t-x)\!(t-y)\!(t-z)}\nonumber\\
&&\hspace{0.5cm}-\frac{n^2\log(n)}{(n-t)\!(n-w)\!(n-x)\!(n-y)\!(n-z)}\Big),\nonumber\\
&&\hspace{4.0cm}I_4(x,y,z,w,t,n)=\frac{\partial{I_3(x,y,z,w,t,n)}}{\partial x},\nonumber\\
&&\hspace{4.0cm}I_5(x,y,z,w,t,n)=\frac{\partial^2{I_3(x,y,z,w,t,n)}}{\partial x\partial y}.
\end{eqnarray}

Here, $x_i=m_i^2/{\Lambda^2}$ with $m_i$ representing the mass of the corresponding particle, and ${\Lambda}$ representing the energy scale of the NP. The coefficient $\mathcal{H}$ is equal to $-\frac{\sqrt{2}}{G_FV_{31}^*V_{33}}$.
\begin{figure}[h]
\setlength{\unitlength}{5.0mm}
\centering
\includegraphics[width=5.0in]{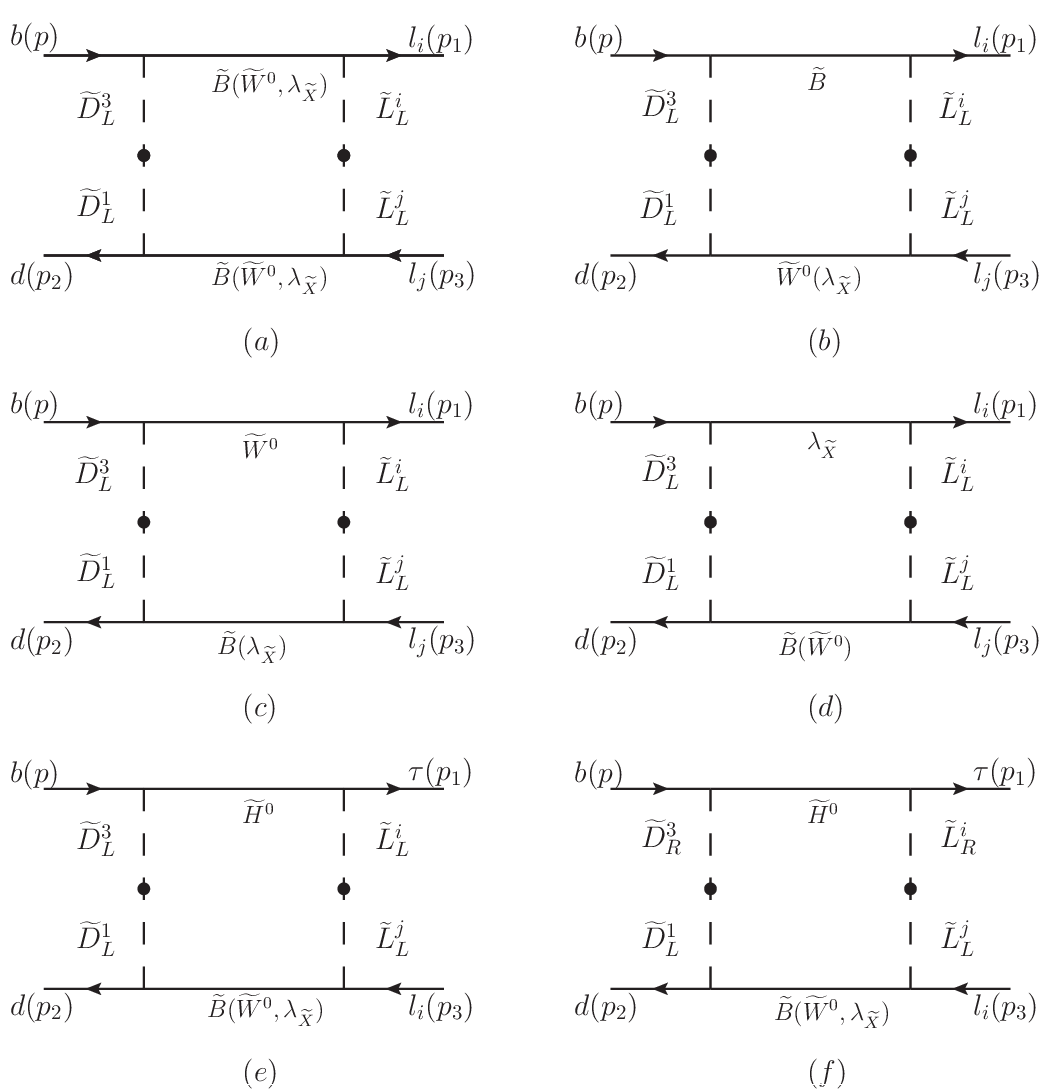}
\caption{The box-type diagrams for the $B^0\rightarrow{{l_i}^{\pm}{l_j}^{\mp}}$ processes of $\chi^0$ in the MIA.}\label{N2}
\end{figure}

The box-type diagrams drawn in Fig.2 can be written as
\begin{eqnarray}
&&T_{box}^2=(B_2^{11}+B_2^{12}+B_2^{13}+B_2^{21}+B_2^{22}+B_2^{31}+B_2^{32}\nonumber\\
&&\hspace{1.2cm}+B_2^{41}+B_2^{42})\overline{l}_i(p_1)\gamma^{\mu}P_Ll_j(p_3)\overline{b}(p)\gamma_{\mu}P_Ld(p_2)\nonumber\\
&&\hspace{1.2cm}+(B_2^{51}+B_2^{52}+B_2^{53})\overline{\tau}(p_1)P_Ll_i(p_3)\overline{b}(p)P_Rd(p_2)\nonumber\\
&&\hspace{1.2cm}+(B_2^{61}+B_2^{62}+B_2^{63})\overline{\tau}(p_1)\gamma^{\mu}P_Ll_i(p_3)\overline{b}(p)\gamma_{\mu}P_Ld(p_2).
\end{eqnarray}

From the box-type diagrams, we obtain the contribution of charged particles to effective couplings $B_2^{\alpha\beta}$ with $\alpha=1...6$ and $\beta=1...3$. For Fig.2(a):
\begin{eqnarray}
&&B_2^{11}=\frac{1}{144\mathcal{H}\Lambda^6}g_1^4\Delta^{LL}_{31}(\widetilde{D})\Delta^{LL}_{ij}(\widetilde{L})
I_2(x_{\widetilde{B}},x_{\widetilde{D}^3_L},x_{\widetilde{D}^1_L},x_{\widetilde{L}^i_L},x_{\widetilde{L}^j_L}),\nonumber\\
&&B_2^{12}=\frac{1}{16\mathcal{H}\Lambda^6}g_2^4\Delta^{LL}_{31}(\widetilde{D})\Delta^{LL}_{ij}(\widetilde{L})
I_2(x_2,x_{\widetilde{D}^3_L},x_{\widetilde{D}^1_L},x_{\widetilde{L}^i_L},x_{\widetilde{L}^j_L}),\nonumber\\
&&B_2^{13}=\frac{1}{144\mathcal{H}\Lambda^6}g_{YX}^4\Delta^{LL}_{31}(\widetilde{D})\Delta^{LL}_{ij}(\widetilde{L})
I_2(x_{\lambda_{\widetilde{X}}},x_{\widetilde{D}^3_L},x_{\widetilde{D}^1_L},x_{\widetilde{L}^i_L},x_{\widetilde{L}^j_L}).
\end{eqnarray}
For Fig.2(b), Fig.2(c) and Fig.2(d):
\begin{eqnarray}
&&B_2^{21}=B_2^{31}=-\frac{1}{48\mathcal{H}\Lambda^6}g_1^2g_2^2\Delta^{LL}_{31}(\widetilde{D})\Delta^{LL}_{ij}(\widetilde{L})
I_3(x_{\widetilde{B}},x_2,x_{\widetilde{D}^3_L},x_{\widetilde{D}^1_L},x_{\widetilde{L}^i_L},x_{\widetilde{L}^j_L}),\nonumber\\
&&B_2^{22}=B_2^{41}=\frac{1}{144\mathcal{H}\Lambda^6}g_1^2g_{YX}^2\Delta^{LL}_{31}(\widetilde{D})\Delta^{LL}_{ij}(\widetilde{L})
I_3(x_{\widetilde{B}},x_{\lambda_{\widetilde{X}}},x_{\widetilde{D}^3_L},x_{\widetilde{D}^1_L},x_{\widetilde{L}^i_L},x_{\widetilde{L}^j_L}),\nonumber\\
&&B_2^{32}=B_2^{42}=-\frac{1}{48\mathcal{H}\Lambda^6}g_2^2g_{YX}^2\Delta^{LL}_{31}(\widetilde{D})\Delta^{LL}_{ij}(\widetilde{L})
I_3(x_2,x_{\lambda_{\widetilde{X}}},x_{\widetilde{D}^3_L},x_{\widetilde{D}^1_L},x_{\widetilde{L}^i_L},x_{\widetilde{L}^j_L}).
\end{eqnarray}
For Fig.2(e):
\begin{eqnarray}
&&B_2^{51}=-\frac{1}{6\mathcal{H}\Lambda^6}\frac{g_1^2m_bm_{\tau}}{v^2\cos^2\beta}\Delta^{LL}_{31}(\widetilde{D})\Delta^{LL}_{ij}(\widetilde{L})
I_3(x_{\mu_H^{'}},x_{\widetilde{B}},x_{\widetilde{D}^3_L},x_{\widetilde{D}^1_L},x_{\widetilde{L}^i_L},x_{\widetilde{L}^j_L}),\nonumber\\
&&B_2^{52}=\frac{1}{2\mathcal{H}\Lambda^6}\frac{g_2^2m_bm_{\tau}}{v^2\cos^2\beta}\Delta^{LL}_{31}(\widetilde{D})\Delta^{LL}_{ij}(\widetilde{L})
I_3(x_{\mu_H^{'}},x_2,x_{\widetilde{D}^3_L},x_{\widetilde{D}^1_L},x_{\widetilde{L}^i_L},x_{\widetilde{L}^j_L}),\nonumber\\
&&B_2^{53}=-\frac{1}{6\mathcal{H}\Lambda^6}\frac{g_{YX}^2m_bm_{\tau}}{v^2\cos^2\beta}\Delta^{LL}_{31}(\widetilde{D})\Delta^{LL}_{ij}(\widetilde{L})
I_3(x_{\mu_H^{'}},x_{\lambda_{\widetilde{X}}},x_{\widetilde{D}^3_L},x_{\widetilde{D}^1_L},x_{\widetilde{L}^i_L},x_{\widetilde{L}^j_L}).
\end{eqnarray}
For Fig.2(f):
\begin{eqnarray}
&&B_2^{61}=-\frac{1}{12\mathcal{H}\Lambda^6}\frac{g_1^2m_bm_{\tau}}{v^2\cos^2\beta}\Delta^{RL}_{31}(\widetilde{D})\Delta^{RL}_{ij}(\widetilde{L})
I_3(x_{\mu_H^{'}},x_{\widetilde{B}},x_{\widetilde{D}^3_R},x_{\widetilde{D}^1_L},x_{\widetilde{L}^i_R},x_{\widetilde{L}^j_L}),\nonumber\\
&&B_2^{62}=\frac{1}{4\mathcal{H}\Lambda^6}\frac{g_2^2m_bm_{\tau}}{v^2\cos^2\beta}\Delta^{RL}_{31}(\widetilde{D})\Delta^{RL}_{ij}(\widetilde{L})
I_3(x_{\mu_H^{'}},x_2,x_{\widetilde{D}^3_R},x_{\widetilde{D}^1_L},x_{\widetilde{L}^i_R},x_{\widetilde{L}^j_L}),\nonumber\\
&&B_2^{63}=-\frac{1}{12\mathcal{H}\Lambda^6}\frac{g_{YX}^2m_bm_{\tau}}{v^2\cos^2\beta}\Delta^{RL}_{31}(\widetilde{D})\Delta^{RL}_{ij}(\widetilde{L})
I_3(x_{\mu_H^{'}},x_{\lambda_{\widetilde{X}}},x_{\widetilde{D}^3_R},x_{\widetilde{D}^1_L},x_{\widetilde{L}^i_R},x_{\widetilde{L}^j_L}).
\end{eqnarray}

\begin{figure}[h]
\setlength{\unitlength}{5.0mm}
\centering
\includegraphics[width=5.0in]{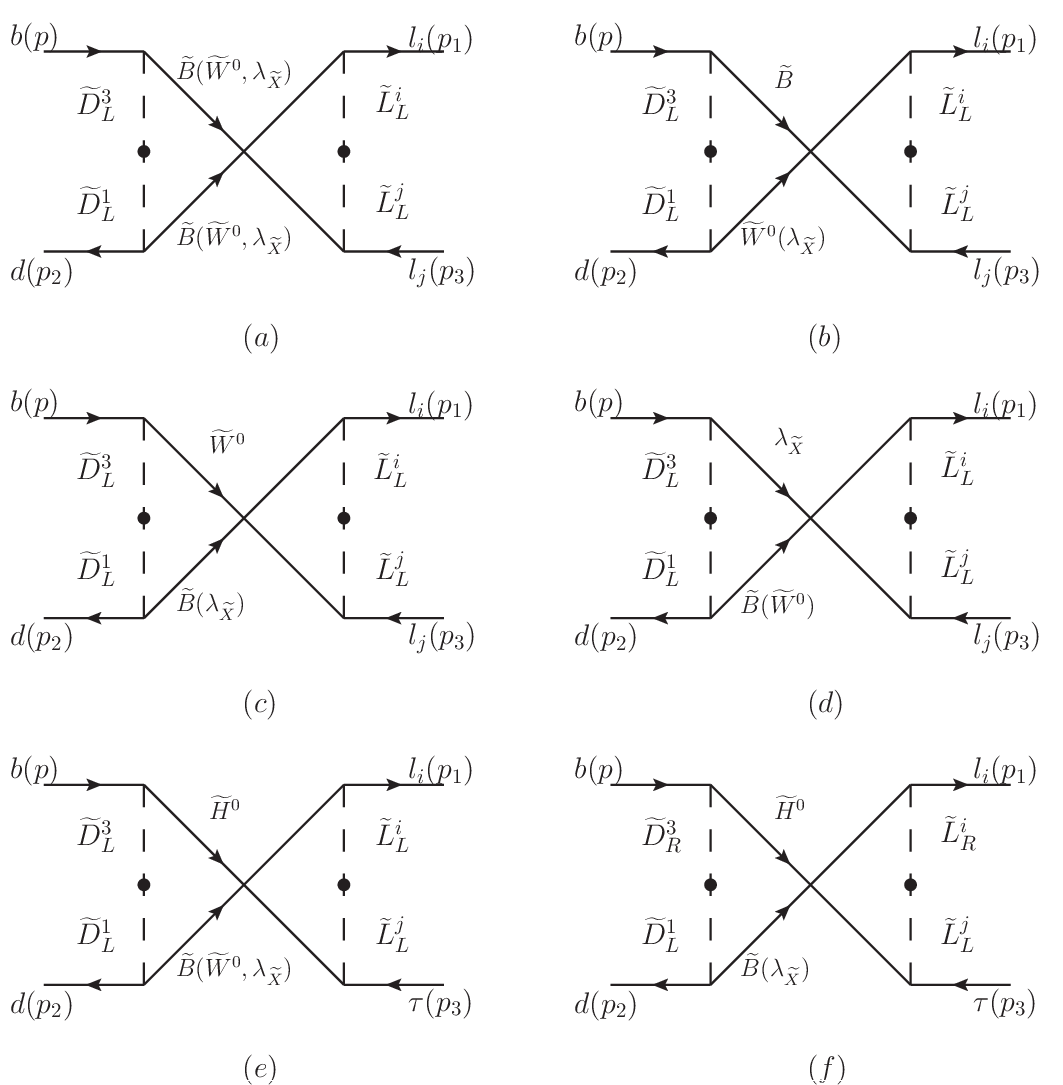}
\caption{The box-type diagrams for the $B^0\rightarrow{{l_i}^{\pm}{l_j}^{\mp}}$ processes of $\chi^0$ in the MIA.}\label{N3}
\end{figure}

The box-type diagrams drawn in Fig.3 can be written as
\begin{eqnarray}
&&T_{box}^3=(B_3^{11}+B_3^{12}+B_3^{13}+B_3^{21}+B_3^{22}+B_3^{31}+B_3^{32}\nonumber\\
&&\hspace{1.2cm}+B_3^{41}+B_3^{42})\overline{l}_i(p_1)\gamma^{\mu}P_Ll_j(p_3)\overline{b}(p)\gamma_{\mu}P_Ld(p_2)\nonumber\\
&&\hspace{1.2cm}+(B_3^{51}+B_3^{52}+B_3^{53})\overline{l}_i(p_1)P_R\tau(p_3)\overline{b}(p)P_Ld(p_2)\nonumber\\
&&\hspace{1.2cm}+(B_3^{61}+B_3^{62})\overline{l}_i(p_1)\gamma^{\mu}P_L\tau(p_3)\overline{b}(p)\gamma_{\mu}P_Ld(p_2).
\end{eqnarray}

From the box-type diagrams, we obtain the contribution of charged particles to effective couplings $B_3^{\alpha\beta}$ with $\alpha=1...6$ and $\beta=1...3$. For Fig.3(a):
\begin{eqnarray}
&&B_3^{11}=-\frac{1}{144\mathcal{H}\Lambda^6}g_1^4\Delta^{LL}_{31}(\widetilde{D})\Delta^{LL}_{ij}(\widetilde{L})
I_2(x_{\widetilde{B}},x_{\widetilde{D}^3_L},x_{\widetilde{D}^1_L},x_{\widetilde{L}^i_L},x_{\widetilde{L}^j_L}),\nonumber\\
&&B_3^{12}=-\frac{1}{16\mathcal{H}\Lambda^6}g_2^4\Delta^{LL}_{31}(\widetilde{D})\Delta^{LL}_{ij}(\widetilde{L})
I_2(x_2,x_{\widetilde{D}^3_L},x_{\widetilde{D}^1_L},x_{\widetilde{L}^i_L},x_{\widetilde{L}^j_L}),\nonumber\\
&&B_3^{13}=-\frac{1}{144\mathcal{H}\Lambda^6}g_{YX}^4\Delta^{LL}_{31}(\widetilde{D})\Delta^{LL}_{ij}(\widetilde{L})
I_2(x_{\lambda_{\widetilde{X}}},x_{\widetilde{D}^3_L},x_{\widetilde{D}^1_L},x_{\widetilde{L}^i_L},x_{\widetilde{L}^j_L}).
\end{eqnarray}
For Fig.3(b), Fig.3(c) and Fig.3(d):
\begin{eqnarray}
&&B_3^{21}=B_3^{31}=\frac{1}{48\mathcal{H}\Lambda^6}g_1^2g_2^2\Delta^{LL}_{31}(\widetilde{D})\Delta^{LL}_{ij}(\widetilde{L})
I_3(x_{\widetilde{B}},x_2,x_{\widetilde{D}^3_L},x_{\widetilde{D}^1_L},x_{\widetilde{L}^i_L},x_{\widetilde{L}^j_L}),\nonumber\\
&&B_3^{22}=B_3^{41}=-\frac{1}{144\mathcal{H}\Lambda^6}g_1^2g_{YX}^2\Delta^{LL}_{31}(\widetilde{D})\Delta^{LL}_{ij}(\widetilde{L})
I_3(x_{\widetilde{B}},x_{\lambda_{\widetilde{X}}},x_{\widetilde{D}^3_L},x_{\widetilde{D}^1_L},x_{\widetilde{L}^i_L},x_{\widetilde{L}^j_L}),\nonumber\\
&&B_3^{32}=B_3^{42}=\frac{1}{48\mathcal{H}\Lambda^6}g_2^2g_{YX}^2\Delta^{LL}_{31}(\widetilde{D})\Delta^{LL}_{ij}(\widetilde{L})
I_3(x_2,x_{\lambda_{\widetilde{X}}},x_{\widetilde{D}^3_L},x_{\widetilde{D}^1_L},x_{\widetilde{L}^i_L},x_{\widetilde{L}^j_L}).
\end{eqnarray}
For Fig.3(e):
\begin{eqnarray}
&&B_3^{51}=-\frac{1}{6\mathcal{H}\Lambda^6}\frac{g_1^2m_bm_{\tau}}{v^2\cos^2\beta}\Delta^{LL}_{31}(\widetilde{D})\Delta^{LL}_{ij}(\widetilde{L})
I_3(x_{\mu_H^{'}},x_{\widetilde{B}},x_{\widetilde{D}^3_L},x_{\widetilde{D}^1_L},x_{\widetilde{L}^i_L},x_{\widetilde{L}^j_L}),\nonumber\\
&&B_3^{52}=\frac{1}{2\mathcal{H}\Lambda^6}\frac{g_2^2m_bm_{\tau}}{v^2\cos^2\beta}\Delta^{LL}_{31}(\widetilde{D})\Delta^{LL}_{ij}(\widetilde{L})
I_3(x_{\mu_H^{'}},x_2,x_{\widetilde{D}^3_L},x_{\widetilde{D}^1_L},x_{\widetilde{L}^i_L},x_{\widetilde{L}^j_L}),\nonumber\\
&&B_3^{53}=-\frac{1}{6\mathcal{H}\Lambda^6}\frac{g_{YX}^2m_bm_{\tau}}{v^2\cos^2\beta}\Delta^{LL}_{31}(\widetilde{D})\Delta^{LL}_{ij}(\widetilde{L})
I_3(x_{\mu_H^{'}},x_{\lambda_{\widetilde{X}}},x_{\widetilde{D}^3_L},x_{\widetilde{D}^1_L},x_{\widetilde{L}^i_L},x_{\widetilde{L}^j_L}).
\end{eqnarray}
For Fig.3(f):
\begin{eqnarray}
&&B_3^{61}=-\frac{1}{6\mathcal{H}\Lambda^6}\frac{g_1^2m_bm_{\tau}}{v^2\cos^2\beta}\Delta^{RL}_{31}(\widetilde{D})\Delta^{RL}_{ij}(\widetilde{L})
I_3(x_{\widetilde{B}},x_{\mu_H^{'}},x_{\widetilde{D}^3_R},x_{\widetilde{D}^1_L},x_{\widetilde{L}^i_R},x_{\widetilde{L}^j_L}),\nonumber\\
&&B_3^{62}=-\frac{1}{12\mathcal{H}\Lambda^6}\frac{g_{YX}(2g_{YX}+g_X)m_bm_{\tau}}{v^2\cos^2\beta}\Delta^{RL}_{31}(\widetilde{D})\Delta^{RL}_{ij}(\widetilde{L})\nonumber\\
&&\hspace{1.2cm}\times I_3(x_{\lambda_{\widetilde{X}}},x_{\mu_H^{'}},x_{\widetilde{D}^3_R},x_{\widetilde{D}^1_L},x_{\widetilde{L}^i_R},x_{\widetilde{L}^j_L}).
\end{eqnarray}

According to the above content, we can get the coefficients of all graphs. By adding the relevant coefficients, we can obtain the corresponding $C_i(i=1...5)$
\begin{eqnarray}
&&C_1=B_1^1+B_1^3+B_1^4+B_1^5+B_2^{11}+B_2^{12}+B_2^{13}+B_2^{21}\nonumber\\
&&\hspace{1.0cm}+B_2^{22}+B_2^{31}+B_2^{32}+B_2^{41}+B_2^{42}+B_3^{11}+B_3^{12}\nonumber\\
&&\hspace{1.0cm}+B_3^{13}+B_3^{21}+B_3^{22}+B_3^{31}+B_3^{32}+B_3^{41}+B_3^{42},\nonumber\\
&&C_2=B_1^2+B_1^6+B_2^{51}+B_2^{52}+B_2^{53},\hspace{0.5cm}C_5=B_3^{61}+B_3^{62},\nonumber\\
&&C_3=B_2^{61}+B_2^{62}+B_2^{63} ,\hspace{1.5cm}  C_4=B_3^{51}+B_3^{52}+B_3^{53}.
\end{eqnarray}

\subsection{ Degenerate Result}
In this section, we assume that the masses of all superparticles are almost degenerate. In this way, we can more directly analyze the factors affecting the LFV processes $B^0\rightarrow{{l_i}^{\pm}{l_j}^{\mp}}$. We give the one-loop results for the extreme case, where the superparticles (~$M_1,~M_2,~M_{\mu_H^\prime},~M_{\lambda_{\tilde{X}}},~M_{BB^\prime},~M_{\tilde{D}_L},~M_{\tilde{D}_R},
~M_{\tilde{U}_L},~M_{\tilde{U}_R},~M_{\tilde{L}_L},~M_{\tilde{L}_R},~M_{\tilde{\nu}_L},~M_{\tilde{\nu}_R} $~) have the same mass as $M_{SUSY}$:
\begin{eqnarray}
&&|M_1|=|M_2|=|M_{\mu_H^{'}}|=|M_{\lambda_{\widetilde{X}}}|=|M_{BB^{'}}|=M_{\widetilde{D}_L}=M_{\widetilde{D}_R}\nonumber\\
&&=M_{\widetilde{U}_L}=M_{\widetilde{U}_R}=M_{\widetilde{L}_L}=M_{\widetilde{L}_R}=M_{\widetilde{\nu}_L}
=M_{\widetilde{\nu}_R}=M_{SUSY}.
\end{eqnarray}

The functions $I_i(i = 2...5)$, $\Delta^{AB}_{ij(nk)}(C)(A,B = R,L; C=\widetilde{D},\widetilde{U},\widetilde{L},\widetilde{\nu})$, $T_{u}^{nk}$, $T_{e}^{ij}$ and $T_{d}^{31}$ are much simplified as
\begin{eqnarray}
&&I_2(1,1,1,1,1)=-\frac{\rm i}{480\pi^2},\hspace{1.8cm}I_3(1,1,1,1,1,1)=-\frac{\rm i}{480\pi^2},\nonumber\\
&&I_4(1,1,1,1,1,1)=\frac{\rm i}{960\pi^2},\hspace{1.8cm}I_5(1,1,1,1,1,1)=-\frac{\rm i}{1680\pi^2},
%%%%%%%%%%%%%%%%%%%%%%%%%%%%%%%%%%%%%%%%%
\\&&\Delta^{LL}_{nk}(\widetilde{U})=M^2_{SUSY}\delta^{LL}_{nk}(\widetilde{U}),\hspace{1.8cm}
\Delta^{RR}_{nk}(\widetilde{U})=M^2_{SUSY}\delta^{RR}_{nk}(\widetilde{U}),\nonumber\\
&&\Delta^{LL}_{31}(\widetilde{D})=M^2_{SUSY}\delta^{LL}_{31}(\widetilde{D}),\hspace{1.7cm}
\Delta^{RL}_{31}(\widetilde{D})=\frac{\sqrt{2}}{2}v_dT_{d}^{31},\nonumber\\
&&\Delta^{LL}_{ij}(\widetilde{L})=M^2_{SUSY}\delta^{LL}_{ij}(\widetilde{L}),\hspace{1.8cm}
\Delta^{LL}_{ij}(\widetilde{\nu})=M^2_{SUSY}\delta^{LL}_{ij}(\widetilde{\nu}),\nonumber\\
&&\Delta^{LR}_{nk}(\widetilde{U})=\Delta^{RL}_{nk}(\widetilde{U})=\frac{\sqrt{2}}{2}v_uT_{u}^{nk},\hspace{0.6cm}
{\Delta^{RL}_{ij}(\widetilde{L})=\frac{\sqrt{2}}{2}v_dT_{e}^{ij}},\nonumber\\
&&T_{u}^{nk}=M_{SUSY}\delta_{T_u}^{nk},\hspace{0.9cm}T_{d}^{31}=M_{SUSY}\delta_{T_d}^{31},\hspace{0.9cm}
T_{e}^{ij}=M_{SUSY}\delta_{T_e}^{ij}.
\end{eqnarray}

Then, we obtain the much simplified one-loop results of $C_i(i=1...5)$
\begin{eqnarray}
&&C_1=-\frac{{\rm i}g_2^2}{480\pi^2M^2_{SUSY}}\{\frac{1}{8}g_2^2\sum_{n,k=1}^{3}
\delta^{LL}_{nk}(\widetilde{U})\delta^{LL}_{ij}(\widetilde{\nu})V_{3n}V_{1k}^{*}+\frac{m_W}{M^2_{SUSY}}\big[
\frac{m_W}{7v^2\sin^2\beta}\nonumber\\
&&\hspace{1.0cm}\times(\texttt{sign}[M_2 \mu_H^{'}]+2\sin\beta \cos\beta)\sum_{n,k=1}^{3}\delta^{RR}_{nk}(\widetilde{U})
\delta^{LL}_{ij}(\widetilde{\nu})m_{u^n}m_{u^k}V_{3n}V_{1k}^{*}+\frac{\sqrt{2}g_2}{16}\nonumber\\
&&\hspace{1.0cm}\times(\cos\beta \texttt{sign}[\mu_H^{'}]+\sin\beta \texttt{sign}[M_2])\sum_{n,k=1}^{3}
\delta^{nk}_{T_u}\delta^{LL}_{ij}(\widetilde{\nu})m_{u^n}V_{3n}V_{1k}^{*}+\frac{\sqrt{2}g_2}{16}\nonumber\\
&&\hspace{1.0cm}\times(\cos\beta \texttt{sign}[M_2]+\sin\beta \texttt{sign}[\mu_H^{'}])
\sum_{n,k=1}^{3}\delta^{nk}_{T_u}\delta^{LL}_{ij}(\widetilde{\nu})m_{u^k}V_{3n}V_{1k}^{*}\big]\},
%%%%%%%%%%%%%%%%%%%%%%%%%%%%%%%%%%%%%%%%%
\\&&C_2=-\frac{{\rm i}}{960\pi^2M^2_{SUSY}}\frac{m_bm_{\tau}}{v^2\cos^2\beta}\{g_2^2\sum_{n,k=1}^{3}
\delta^{LL}_{nk}(\widetilde{U})\delta^{LL}_{ij}(\widetilde{\nu})V_{3n}V_{1k}^{*}-(\frac{1}{3}g_1^2\nonumber\\
&&\hspace{1.0cm}-g_2^2+\frac{1}{3}g_{YX}^2)\delta^{LL}_{31}(\widetilde{D})\delta^{LL}_{ij}(\widetilde{L})+\frac{\sqrt{2}g_2m_W}{2M^2_{SUSY}}
(\cos\beta \texttt{sign}[M_2]\nonumber\\
&&\hspace{1.0cm}+\sin\beta \texttt{sign}[\mu_H^{'}]))\sum_{n,k=1}^{3}\delta^{nk}_{T_u}\delta^{LL}_{ij}(\widetilde{\nu})m_{u^k}V_{3n}V_{1k}^{*}\},
%%%%%%%%%%%%%%%%%%%%%%%%%%%%%%%%%%%%%%%%%
\\&&C_3=\frac{{\rm i}m_bm_{\tau}}{3840\pi^2M^4_{SUSY}}(\frac{1}{3}g_1^2-g_2^2+\frac{1}{3}g_{YX}^2)\delta^{31}_{T_d}
\delta^{ij}_{T_e},\nonumber\\
&&C_4=\frac{{\rm i}}{960\pi^2M^2_{SUSY}}\frac{m_bm_{\tau}}{v^2\cos^2\beta}(\frac{1}{3}g_1^2-g_2^2+\frac{1}{3}g_{YX}^2)
\delta^{LL}_{31}(\widetilde{D})\delta^{LL}_{ij}(\widetilde{L}),\nonumber\\
&&C_5=\frac{{\rm i}m_bm_{\tau}}{5760\pi^2M^4_{SUSY}}\Big(g_1^2+\frac{1}{2}g_{YX}(2g_{YX}+g_X)\Big)\delta^{31}_{T_d}\delta^{ij}_{T_e}.
\end{eqnarray}

From the results of the above formula, we can find that $\texttt{sign}[M_2]$ and $\texttt{sign}[\mu_H^{'}]$  have a certain impact on the corrections of $C_1$ and $C_2$. According to  $1>g_X>g_{YX}>0$, we assume positive signs for $M_2$ and $\mu_H^{'}$ in order to obtain larger values for $C_1$ and $C_2$
\begin{eqnarray}
&&C_1=-\frac{{\rm i}g_2^2}{480\pi^2M^2_{SUSY}}\{\frac{1}{8}g_2^2\sum_{n,k=1}^{3}
\delta^{LL}_{nk}(\widetilde{U})\delta^{LL}_{ij}(\widetilde{\nu})V_{3n}V_{1k}^{*}+\frac{m_W}{M^2_{SUSY}}\big[\frac{m_W}{7v^2\sin^2\beta}\nonumber\\
&&\hspace{1.0cm}\times (1+2\sin\beta \cos\beta)\!\sum_{n,k=1}^{3}\delta^{RR}_{nk}(\widetilde{U})
\delta^{LL}_{ij}(\widetilde{\nu})m_{u^n}m_{u^k}V_{3n}V_{1k}^{*}+\frac{\sqrt{2}g_2}{16}(\cos\beta+\sin\beta)\nonumber\\
&&\hspace{1.0cm}\times \!\sum_{n,k=1}^{3}
\!\delta^{nk}_{T_u}\delta^{LL}_{ij}(\widetilde{\nu})m_{u^n}V_{3n}V_{1k}^{*}\!+\!\frac{\sqrt{2}g_2}{16}(\cos\beta\!+\!\sin\beta)
\sum_{n,k=1}^{3}\!\delta^{nk}_{T_u}\delta^{LL}_{ij}(\widetilde{\nu})m_{u^k}V_{3n}V_{1k}^{*}\big]\},\nonumber\\
&&C_2=-\frac{{\rm i}}{960\pi^2M^2_{SUSY}}\frac{m_bm_{\tau}}{v^2\cos^2\beta}\{g_2^2\sum_{n,k=1}^{3}
\delta^{LL}_{nk}(\widetilde{U})\delta^{LL}_{ij}(\widetilde{\nu})V_{3n}V_{1k}^{*}-(\frac{1}{3}g_1^2-g_2^2+\frac{1}{3}g_{YX}^2)\nonumber\\
&&\hspace{1.0cm}\times\delta^{LL}_{31}(\widetilde{D})
\delta^{LL}_{ij}(\widetilde{L})+\frac{\sqrt{2}g_2m_W}{2M^2_{SUSY}} (\cos\beta+\sin\beta)\sum_{n,k=1}^{3}\delta^{nk}_{T_u}
\delta^{LL}_{ij}(\widetilde{\nu})m_{u^k}V_{3n}V_{1k}^{*}\}.
\end{eqnarray}

Combined with the decoupling results of the above formula, we set $g_{YX}=0.1,~g_X=0.4,~M_{SUSY}=\mu_H',~\Delta^{RR}_{nk}(\tilde{U})=\Delta^{LL}_{nk}(\tilde{U})=1\times10^{4}~{\rm{GeV}^2}(n\ne k ~{\rm and}~ n,k=1,2,3),~\Delta^{RR}_{nn}(\tilde{U})=\Delta^{LL}_{ii}(\tilde{U})=2.5\times10^{6}~{\rm{GeV}^2}~(i,n=1,2,3),$ and $\Delta^{LL}_{31}(\tilde{D})=1\times10^{4}~{\rm{GeV}^2}~$ and discuss in three cases:

\begin{figure}[h]
\setlength{\unitlength}{5mm}
\centering
\includegraphics[width=3.0in]{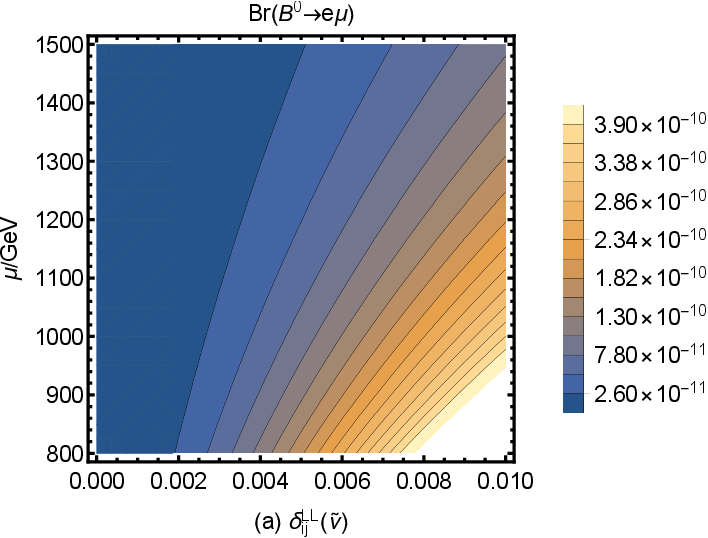}
\vspace{0.2cm}
\setlength{\unitlength}{5mm}
\centering
\includegraphics[width=3.0in]{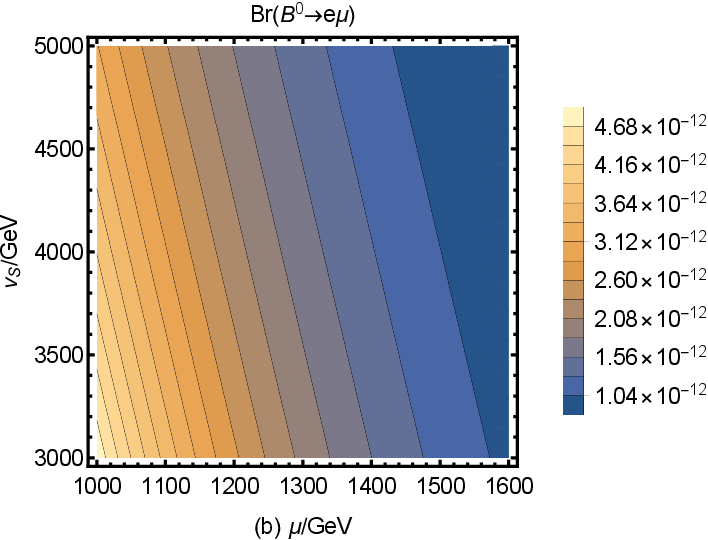}
\vspace{0.2cm}
\setlength{\unitlength}{5mm}
\centering
\includegraphics[width=3.0in]{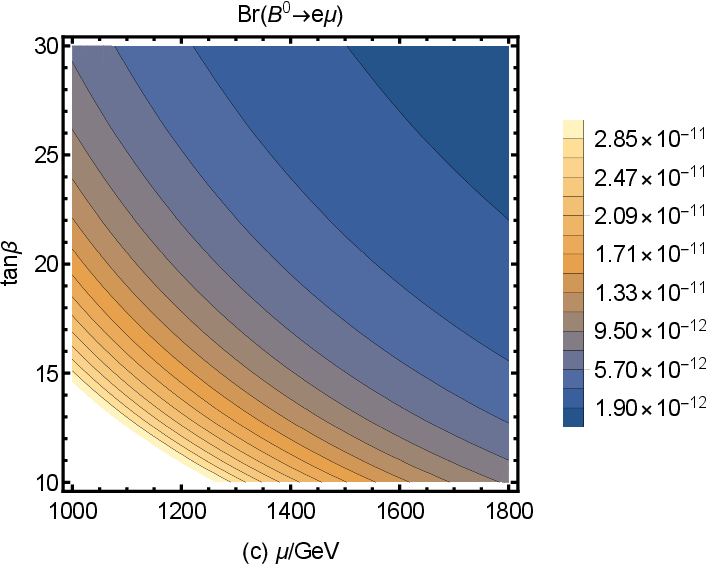}
\caption{ Fig.\ref {N4}(a) shows the effect of $\delta^{LL}_{ij}(\tilde{\nu})$ and $\mu$ on $Br(B^0\rightarrow e{\mu})$ with $\tan\beta=20$ and $v_S=4.3~\rm TeV$. Fig.\ref {N4}(b) shows the effect of $v_S$ and $\mu$ on $Br(B^0\rightarrow e{\mu})$ with $\tan\beta=20$ and $\delta^{LL}_{ij}(\tilde{\nu})=10^{-3}$. Fig.\ref {N4}(c) shows the effect of $\tan\beta$ and $\mu$ on $Br(B^0\rightarrow e{\mu})$ with $\delta^{LL}_{ij}(\tilde{\nu})=2\times10^{-3}$ and $v_S=3.8~\rm TeV$. The right icon represents the color corresponding to the $Br(B^0\rightarrow e{\mu})$ value. }\label{N4}
\end{figure}

1.$B^0\rightarrow e{\mu}$

In order to investigate the parameters that affect the branching ratio of $ B^0\rightarrow{e\mu}$, we take $\tan\beta,~\delta^{LL}_{ij}(\tilde{\nu}),~\mu$ and $v_S$ as the variables. In the Fig.\ref {N4}(a), we set $\tan\beta=20$ and $v_S=4.3~\rm TeV$ to explore the influence of $\delta^{LL}_{ij}(\tilde{\nu})$ and $\mu$ on $Br(B^0\rightarrow e{\mu})$. From the graph we can clearly see that the value of $Br(B^0\rightarrow e{\mu})$ increases with the increase of $\delta^{LL}_{ij}(\tilde{\nu})$ and decreases with the increase of $\mu$. Both $\delta^{LL}_{ij}(\tilde{\nu})$ and $\mu$ have a significant effect on $Br(B^0\rightarrow e{\mu})$, but the effect of $\delta^{LL}_{ij}(\tilde{\nu})$ is greater than the effect of $\mu$ on $Br(B^0\rightarrow e{\mu})$.

In the Fig.\ref {N4}(b), we set $\tan\beta=20$ and $\delta^{LL}_{ij}(\tilde{\nu})=10^{-3}$ to probe the influence of $v_S$ and $\mu$ on $Br(B^0\rightarrow e{\mu})$. From the figure, it can be found that both $v_S$ and $\mu$ have a strong effect on $Br(B^0\rightarrow e{\mu})$ and they both show a decreasing trend. That is, the smaller the values of $v_S$ and $\mu$, the easier it is to approach the experimental upper limit of the branching ratio. In addition, we can discover that the effect of $v_S$ on the branching ratio is weaker than the effect of $\mu$.

In Fig.\ref {N4}(a) and Fig.\ref {N4}(b), we set $\tan\beta=20$ and $v_S=4.3~\rm TeV$ (and $\delta_{ij}^{LL} =10^{-3}$). For the selection of values in these graphs, we are taking values within the reasonable value space of these parameters. A particular value is chosen because it is the set of values that best reflects the pattern of change after a lot of numerical exploration. Through the analysis of different values, we can find that they have little influence on the overall law of these processes.

In addition to the above research, we also explore the effect of $\tan\beta$ and $\mu$ on $Br(B^0\rightarrow e{\mu})$ with $\delta^{LL}_{ij}(\tilde{\nu})=2\times10^{-3}$ and $v_S=3.8~\rm TeV$ in the Fig.\ref {N4}(c). Except for $\mu$, the effect of $\tan\beta$ on $Br(B^0\rightarrow e{\mu})$ is also great, and $Br(B^0\rightarrow e{\mu})$ decreases as $\tan\beta$ increases. The smaller the value of $\tan\beta$ the closer the experimental upper limit of $Br(B^0\rightarrow e{\mu})$ can be.

2.$B^0\rightarrow e{\tau}$

\begin{figure}[h]
\setlength{\unitlength}{5mm}
\centering
\includegraphics[width=3.0in]{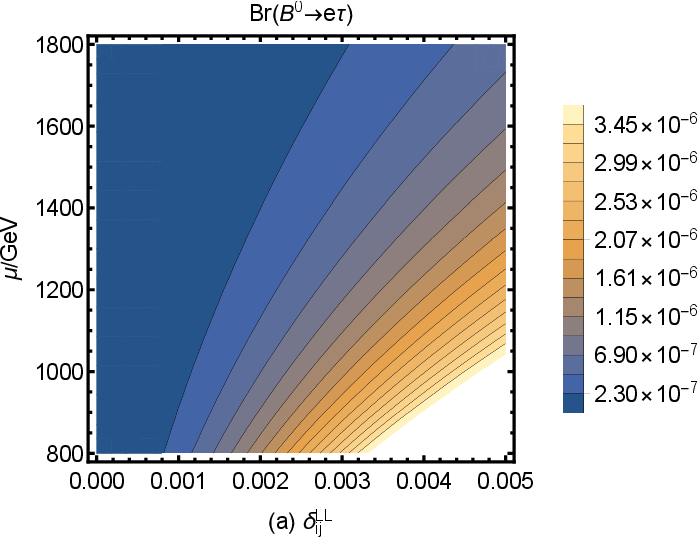}
\vspace{0.2cm}
\setlength{\unitlength}{5mm}
\centering
\includegraphics[width=2.85in]{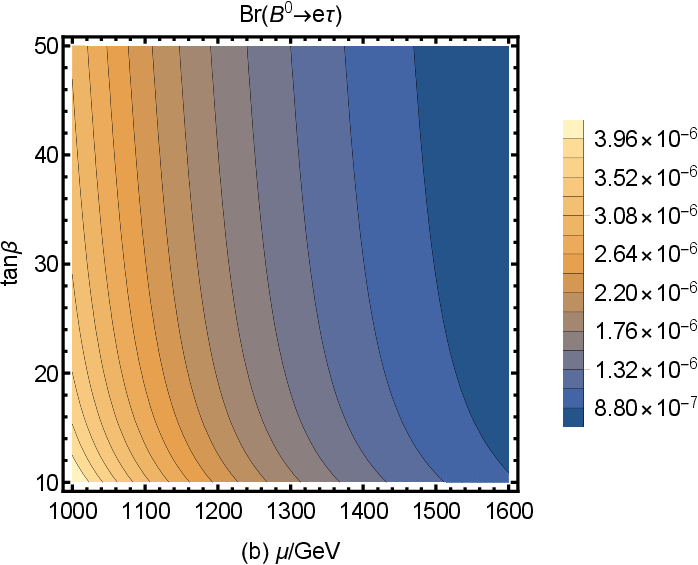}
\vspace{0.2cm}
\setlength{\unitlength}{5mm}
\centering
\includegraphics[width=3.0in]{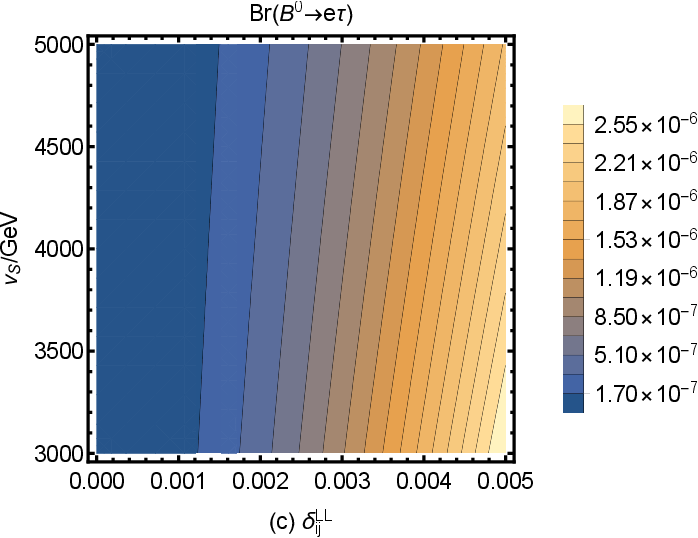}
\caption{ Fig.\ref {N5}(a) shows the effect of $\delta^{LL}_{ij}$ and $\mu$ on $Br(B^0\rightarrow e{\tau})$ with $\tan\beta=10$, $g_{YX}=0.1$ and $v_S=3.8~\rm TeV$. Fig.\ref {N5}(b) shows the effect of $\tan\beta$ and $\mu$ on $Br(B^0\rightarrow e{\tau})$ with $\delta^{LL}_{ij} = 5\times10^{-3}$, $g_{YX}=0.1$ and $v_S=3.8~\rm TeV$. Fig.\ref {N5}(c) shows the effect of $\delta^{LL}_{ij}$ and $v_S$ on $Br(B^0\rightarrow e{\tau})$ with $\tan\beta = 10$, $g_{YX}=0.1$ and $\mu=1.2~\rm TeV$. The right icon represents the color corresponding to the $Br(B^0\rightarrow e{\tau})$ value.}\label{N5}
\end{figure}

For the process of $B^0\rightarrow e{\tau}$, we take $\tan\beta,~\delta^{LL}_{ij},~g_{YX},~\mu$ and $v_S$ as the variables. Here $\delta^{LL}_{ij}$ stands for $\delta^{LL}_{ij}(\tilde{\nu})$ and $\delta^{LL}_{ij}(\tilde{L})$. We set $\tan\beta=10$, $g_{YX}=0.1$ and $v_S=3.8~\rm TeV$ to explore the influence of $\delta^{LL}_{ij}$ and $\mu$ on $Br(B^0\rightarrow e{\tau})$ in the Fig.\ref {N5}(a). We observe that the effect of $\delta^{LL}_{ij}$ and $\mu$ on the branching ratio of the $B^0\rightarrow e{\tau}$ process is also very large. The value of $Br(B^0\rightarrow e{\tau})$ increases as $\delta^{LL}_{ij}$ increases. That is to say, the larger value of $\delta^{LL}_{ij}$ the closer to the experimental limit. The value of $Br(B^0\rightarrow e{\tau})$ decreases as $\mu$ increases, where the smaller value of $\mu$ the closer to the experimental limit. Compared with the change of $\mu$, the change of $\delta^{LL}_{ij}$ has a greater impact on the value of $Br(B^0\rightarrow e{\tau})$.

In the Fig.\ref {N5}(b), we set $\delta^{LL}_{ij} = 5\times10^{-3}$, $g_{YX}=0.1$ and $v_S=3.8~\rm TeV$ to explore the influence of $\tan\beta$ and $\mu$ on $Br(B^0\rightarrow e{\tau})$.
We can observe that $Br(B^0\rightarrow e{\tau})$ decreases with the increase of $\tan\beta$ and $\mu$. Compared $\tan\beta$ with $\mu$, we can find that the influence of parameter $\mu$ on $Br(B^0\rightarrow e{\tau})$ is greater than that of parameter $\tan\beta$ on $Br(B^0\rightarrow e{\tau})$. In the Fig.\ref {N5}(c), we set $\tan\beta = 10$, $g_{YX}=0.1$ and $\mu=1.2~\rm TeV$ to explore the influence of $\delta^{LL}_{ij}$ and $v_S$ on $Br(B^0\rightarrow e{\tau})$. We can find that $Br(B^0\rightarrow e{\tau})$ decreases with the increase of $v_S$, and $Br(B^0\rightarrow e{\tau})$ increase with the increase of $\delta^{LL}_{ij}$. In brief, we observe that $\tan\beta,~\delta^{LL}_{ij},~\mu$ and $v_S$ contribute significantly to the branching ratio of process $B^0\rightarrow e{\tau}$.

3.$B^0\rightarrow {\mu}{\tau}$

\begin{figure}[h]
\setlength{\unitlength}{5mm}
\centering
\includegraphics[width=3.0in]{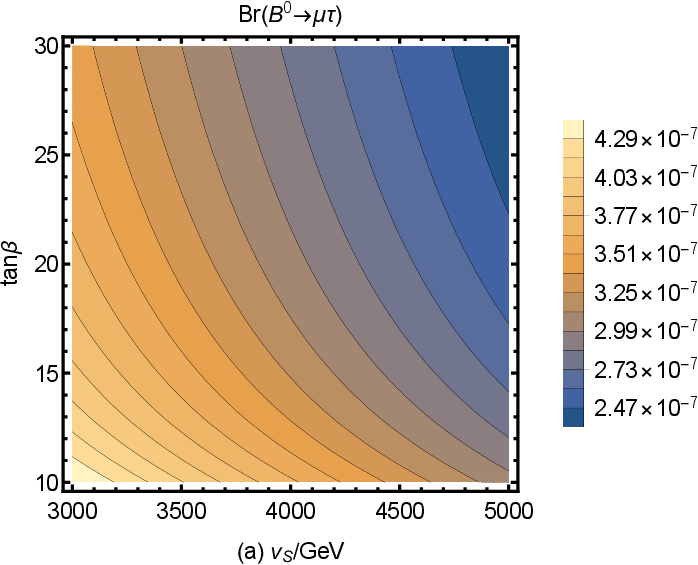}
\vspace{0.2cm}
\setlength{\unitlength}{5mm}
\centering
\includegraphics[width=3.1in]{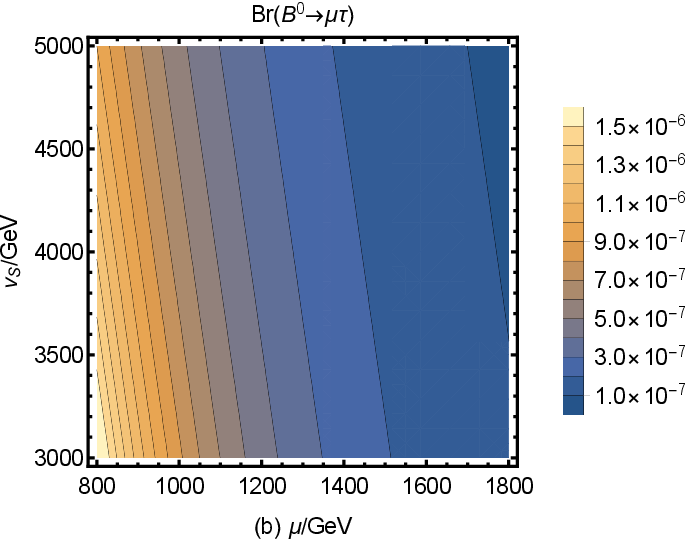}
\caption{ Fig.\ref {N6}(a) shows the effect of $\tan\beta$ and $v_S$ on $Br(B^0\rightarrow {\mu}{\tau})$ with $\delta^{LL}_{ij} = 2\times10^{-3}$, $g_{YX}=0.1$ and $\mu=1.2~\rm TeV$. Fig.\ref {N6}(b) shows the effect of $\mu$ and $v_S$ on $Br(B^0\rightarrow {\mu}{\tau})$ with $\delta^{LL}_{ij} = 2\times10^{-3}$, $g_{YX}=0.1$ and $\tan\beta=10$. The right icon represents the color corresponding to the $Br(B^0\rightarrow {\mu}{\tau})$ value.}\label{N6}
\end{figure}

It's similar to process $B^0\rightarrow e{\tau}$, and we get the Fig.\ref {N6}(a) and Fig.\ref {N6}(b). In the Fig.\ref {N6}(a), we analyze $\tan\beta$ and $v_S$ on $Br(B^0\rightarrow {\mu}{\tau})$. The value of $Br(B^0\rightarrow {\mu}{\tau})$ decreases as the values of both $\tan\beta$ and $v_S$ increases, in which $v_S$ has slightly more effect on $Br(B^0\rightarrow {\mu}{\tau})$ than $\tan\beta$. In the Fig.\ref {N6}(b), we can find that the value of $Br(B^0\rightarrow {\mu}{\tau})$ also decreases as the value of $\mu$ increases. In addition, the influence of $\mu$ on $Br(B^0\rightarrow {\mu}{\tau})$ is more obvious than that of $v_S$.

In conclusion, we can observe that $\tan\beta,~\delta^{LL}_{ij},~\mu$ and $v_S$ have significant contributions to $B^0\rightarrow e{\mu}$, $B^0\rightarrow e{\tau}$ and $B^0\rightarrow {\mu}{\tau}$. We also find that the obtained values of the three processes could well satisfy the experimental limitation.

\section{Numerical analysis}
 In this section, we study the numerical results and consider the constraints from LFV processes
$B^0\rightarrow{{l_i}^{\pm}{l_j}^{\mp}}$.  In order to obtain reasonable numerical results, we need to
study some sensitive parameters, and discuss the processes of $B^0\rightarrow
e\mu$,~$B^0\rightarrow e\tau$,~$B^0\rightarrow \mu\tau$ in three subsections explicitly. We draw the
relation diagrams with different parameters. After analyzing these graphs and the experimental limits of the branching ratios, reasonable parameter spaces are found to explain LFV.

Here, we need to consider the effect of $l_j\rightarrow{l_i\gamma}$ on LFV. The limitation of
$\mu\rightarrow{e\gamma}$ is the strongest, and other restrictions can be achieved if the limit of
$\mu\rightarrow{e\gamma}$ is satisfied\cite{T1}. We also need to consider the constraints imposed on
the NP contribution of the muon anomalous magnetic dipole moment (MDM) $a_{\mu}=(g-2)_{\mu}$ in the
$U(1)_X$SSM. The new averaged experiment value of muon anomaly is
$a_{\mu}^{exp}=116592061(41)\times 10^{-11}(0.35ppm)$ from the latest experimental results\cite{B1}. The
deviation between the experimental measured value of $a_{\mu}$ and the predicted value of SM is given
by\cite{B2,B3} $\Delta a_{\mu}=a_{\mu}^{exp}-a_{\mu}^{SM}=(2.49 \pm 0.48) \times 10^{-9}$, whose
significance is equivalent to 5$\sigma$ level. In our previous work\cite{T10}, we have explored the anomalous magnetic moment of the muon, and the value of $a_{\mu}$ can better meet the experimental limitations.

 According to the latest LHC data\cite{w1,w2,w3,w4,w5}, we need to meet the following conditions: the lightest CP-even Higgs mass $m_{h^0}$=125.25 GeV\cite{B25}; the scalar lepton mass greater than $700~{\rm GeV}$; the chargino mass greater than $1100~{\rm GeV}$; the scalar quark mass greater than $1500~{\rm GeV}$. The latest experimental constraint on the mass of the added heavy vector boson $Z^\prime $ is $M_{Z^{\prime}}> 5.1$ TeV\cite{xin1}. The references\cite{ZPG1,ZPG2} give the upper bound of the ratio of $Z^\prime$ mass to its gauge coupling $M_{Z^\prime}/g_X\geq6$ TeV under 99\% CL.. Taking into account the constraint from LHC data, $\tan \beta_\eta<1.5$\cite{TanBP}. Combined with the above experimental requirements, we obtain a wealth of data, and use graphics to analyze and process these data. Considering the above constraints in the front paragraph, we use the following parameters
\begin{eqnarray}
&&g_X=0.4,~\lambda_H = 0.1,~T_{\lambda_H}=T_{\lambda_C}=T_{\kappa}=1~{\rm TeV},~\kappa=0.1,\nonumber\\
&&M_{BB^\prime}=1.3~{\rm TeV},~M_1=1.1~{\rm TeV},~M_2=1.3~{\rm TeV},~M_{\lambda_{\tilde{X}}}=1.5~{\rm TeV},\nonumber\\
&&l_W = B_{\mu} =B_S=0.1~{\rm TeV}^2,~T_{d}^{31}= 0.1~{\rm TeV},T_{u}^{nn}= 3~{\rm TeV},~(n=1,2,3).
\end{eqnarray}

In order to simplify the numerical study, in the following numerical analysis, we make use of the relationship between parameters, and the parameters vary between them
\begin{eqnarray}
&&\hspace{1.3cm}M^2_{\tilde{L}ij}=M^2_{\tilde{L}ji},~~M^2_{\tilde{\nu}ij}=M^2_{\tilde{\nu}ji},
~~T_{eij}=T_{eji},~~T_{{\nu}ij}=T_{{\nu}ji}(i\ne j).
\end{eqnarray}

Unless we say otherwise, the the non-diagonal elements of the mass matrices are set to zero.

\subsection{$B^0\rightarrow e{\mu}$}
\begin{figure}[h]
\setlength{\unitlength}{5mm}
\centering
\includegraphics[width=3.0in]{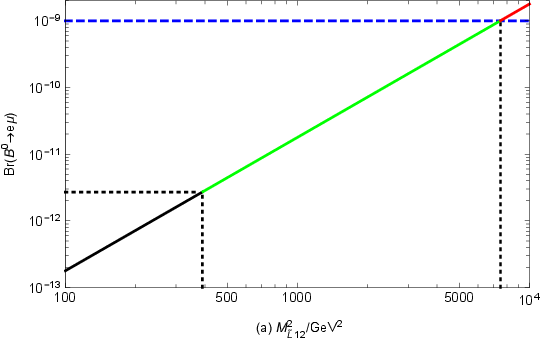}
\vspace{0.2cm}
\setlength{\unitlength}{5mm}
\centering
\includegraphics[width=3.0in]{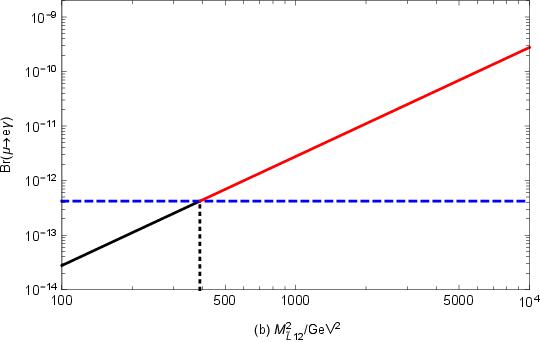}
\vspace{0.2cm}
\setlength{\unitlength}{5mm}
\centering
\includegraphics[width=3.0in]{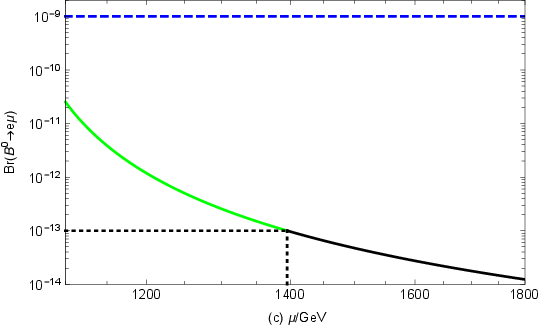}
\vspace{0.2cm}
\setlength{\unitlength}{5mm}
\centering
\includegraphics[width=3.0in]{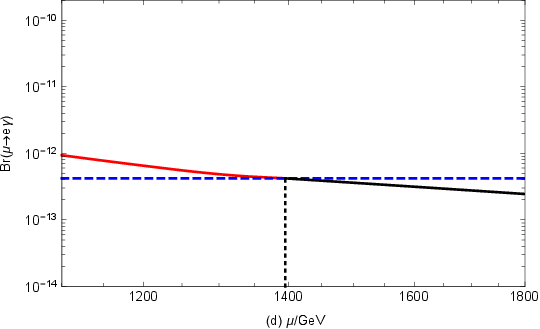}
\vspace{0.2cm}
\setlength{\unitlength}{5mm}
\centering
\includegraphics[width=3.0in]{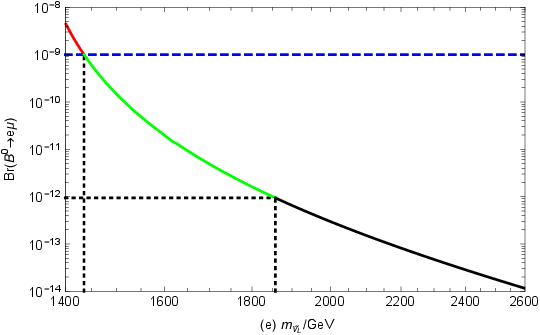}
\vspace{0.2cm}
\setlength{\unitlength}{5mm}
\centering
\includegraphics[width=3.0in]{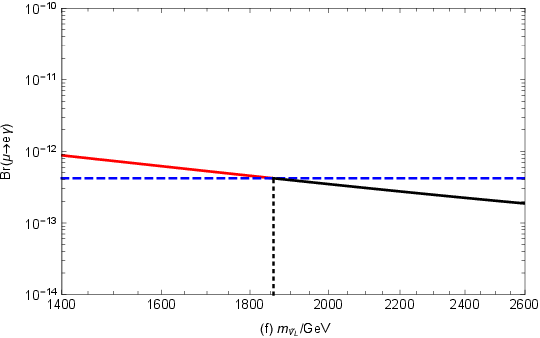}
\caption{ The  diagrams of the effects of $M^2_{\tilde{L}12}$, $\mu$ and $m_{\tilde{\nu}_L}$ on $Br(B^0\rightarrow e{\mu})$ and $Br({\mu}\rightarrow{e\gamma})$. The blue dashed line stands for the upper limit on $Br(B^0\rightarrow e{\mu})$ and $Br({\mu}\rightarrow{e\gamma})$ at 90\%CL.. }\label{N12}
\end{figure}

In order to study the parameters affecting LFV, some sensitive parameters need to be researched. To clearly show the numerical results, with the parameters $v_S=3.8~{\rm TeV}$,~$M^2_{\tilde{L}ii}=1~{\rm TeV}^2$,~$M^2_{{\tilde{\nu}}ii}=0.6~{\rm TeV}^2$,~$M^2_{\tilde{U}nn}=2.6~{\rm TeV}^2$,~(i,n=1,2,3), we plot the trend of $Br(B^0\rightarrow e{\mu})$ and $Br({\mu}\rightarrow{e\gamma})$ affected by different parameters. These diagrams are shown in the Fig.\ref {N12}.

In the figure, the latest experimental upper limits of the branching ratios of $B^0\rightarrow e{\mu}$ and ${\mu}\rightarrow{e\gamma}$ are displayed as the blue dashed line. In the Fig.\ref {N12}(a)(c)(e), the red solid line is excluded by the current limit of $Br(B^0\rightarrow e{\mu})$, and the green solid line indicates that it is consistent with the current limit of $Br(B^0\rightarrow e{\mu})$ but excluded by the current limit of $Br({\mu}\rightarrow{e\gamma})$. The black solid line is consistent with the current limit of $Br({\mu}\rightarrow{e\gamma})$. In the Fig.\ref {N12}(b)(d)(f), the red solid line is excluded by the current limit of $Br({\mu}\rightarrow{e\gamma})$.

First, we consider the off-diagonal terms for the soft breaking slepton mass matrix $M^2_{\tilde{L}}$. In the Fig.\ref {N12}(a), we plot $Br(B^0\rightarrow e{\mu})$ versus $M^2_{\tilde{L}12}$. The line shows an upward trend in  $M^2_{\tilde{L}12}$ range of $100-10^{4}~{\rm GeV}^2$,  which means that $Br(B^0\rightarrow e{\mu})$ increases as $M^2_{\tilde{L}12}$ increases. And $Br(B^0\rightarrow e{\mu})$ will reach the experimental limit with the increase of $M^2_{\tilde{L}12}$. In the Fig.\ref {N12}(b), with the increasing of parameter $M^2_{\tilde{L}12}$, $Br({\mu}\rightarrow{e\gamma})$ will grow rapidly, and $Br({\mu}\rightarrow{e\gamma})$ will soon exceed the experimental limit. As we can see from Fig.\ref {N12}(a)(b), the influences of parameter $M^2_{\tilde{L}12}$ on $Br(B^0\rightarrow e{\mu})$ and $Br({\mu}\rightarrow{e\gamma})$ are huge. The non-diagonal element $M^2_{\tilde{L}ij}(i\ne j)$ leads to strong mixing for sneutrinos and sleptons of different generations. Therefore, the nonzero $M^2_{\tilde{L}ij}(i\ne j)$ enhances LFV and leads to large results.

With the parameters $\tan\beta=5$, $M^2_{\tilde{L}12}=300~{\rm GeV}^2$ and $m_{\tilde{\nu}_L}=800~{\rm GeV}$, Fig.\ref {N12}(c) and Fig.\ref {N12}(d) show that $Br(B^0\rightarrow e{\mu})$ and $Br({\mu}\rightarrow{e\gamma})$ decrease with the increasing of $\mu$. That is to say, the smaller the $\mu$, the greater the value of $Br(B^0\rightarrow e{\mu})$ and $Br({\mu}\rightarrow{e\gamma})$.
With the parameter $\mu_H^{'}=\mu+\frac{\lambda_H v_S}{\sqrt{2}}$, as $\mu$ increases the particle mass increases,
which suppresses LFV significantly. $Br({\mu}\rightarrow{e\gamma})$ can reach the experimental upper limit but $Br(B^0\rightarrow e{\mu})$ cannot do that. As can be seen from the figure, the influence of $\mu$ on $Br(B^0\rightarrow e{\mu})$ is more obvious than that of $Br({\mu}\rightarrow{e\gamma})$.

In addition, we show $Br(B^0\rightarrow e{\mu})$ and $Br({\mu}\rightarrow{e\gamma})$ varying with $m_{\tilde{\nu}_L}$ in the Fig.\ref {N12}(e) and Fig.\ref {N12}(f). We set parameters $\tan\beta=20$, $M^2_{\tilde{L}12}=500~{\rm GeV}^2$ and $\mu=1200~{\rm GeV}$. When the parameter $m_{\tilde{\nu}_L}$ increases, $Br(B^0\rightarrow e{\mu})$ and $Br({\mu}\rightarrow{e\gamma})$ will decrease. Both $Br(B^0\rightarrow e{\mu})$ and $Br({\mu}\rightarrow{e\gamma})$ can reach the experimental upper limit. As can be seen from the figure, the influence of $m_{\tilde{\nu}_L}$ on $Br(B^0\rightarrow e{\mu})$ is more obvious than that of $Br({\mu}\rightarrow{e\gamma})$.

We can briefly analyze the influences of $M^2_{\tilde{L}12}$, $\mu$, and $m_{\tilde{\nu}_L}$ on LFV from the above six figures. As the value of $M^2_{\tilde{L}12}$ increases, the effect of LFV is amplified. The value of $M^2_{\tilde{L}12}$ has a greater influence on the process. For parameters $\mu$ and $m_{\tilde{\nu}_L}$, LFV is depressed as their values increase. When we consider the constraint from $Br({\mu}\rightarrow{e\gamma})$ to $Br(B^0\rightarrow e{\mu})$, $Br(B^0\rightarrow e{\mu})$ can be up to $10^{-12}$. Therefore, it can be seen that the limit of $Br({\mu}\rightarrow{e\gamma})$ to $Br(B^0\rightarrow e{\mu})$ is very strict, which makes $Br(B^0\rightarrow e{\mu})$ difficult to reach the experimental upper limit.
\subsection{$B^0\rightarrow e{\tau}$}
\begin{figure}[h]
\setlength{\unitlength}{5mm}
\centering
\includegraphics[width=3.0in]{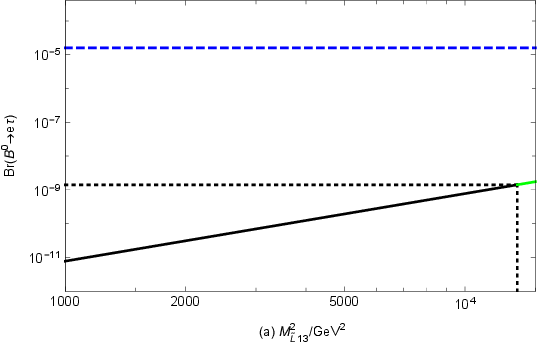}
\vspace{0.2cm}
\setlength{\unitlength}{5mm}
\centering
\includegraphics[width=3.0in]{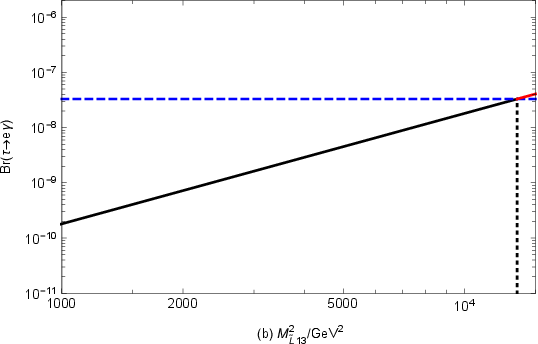}
\vspace{0.2cm}
\setlength{\unitlength}{5mm}
\centering
\includegraphics[width=3.0in]{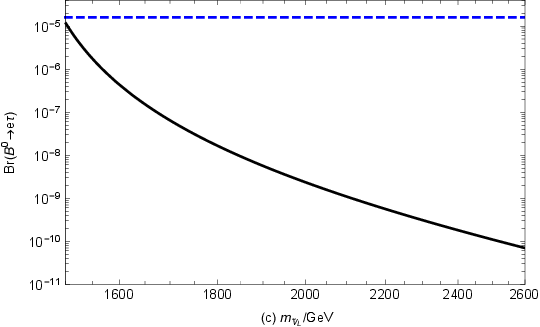}
\vspace{0.2cm}
\setlength{\unitlength}{5mm}
\centering
\includegraphics[width=3.0in]{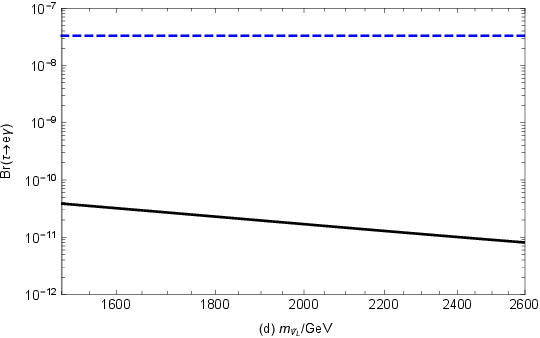}
\vspace{0.2cm}
\setlength{\unitlength}{5mm}
\centering
\includegraphics[width=3.0in]{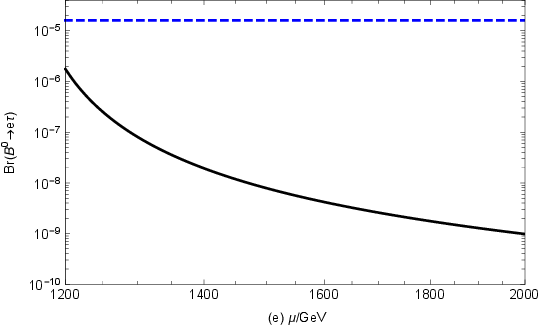}
\vspace{0.2cm}
\setlength{\unitlength}{5mm}
\centering
\includegraphics[width=3.0in]{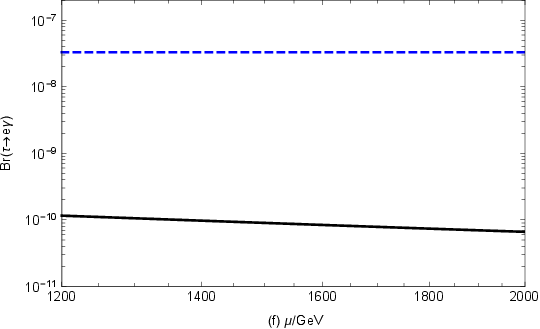}
\caption{ $Br(B^0\rightarrow e{\tau})$ and $Br({\tau}\rightarrow{e\gamma})$ diagrams affected by different parameters. The blue dashed line stands for the upper
limit on $Br(B^0\rightarrow e{\tau})$ and $Br({\tau}\rightarrow{e\gamma})$ at 90\%CL.. }\label{N13}
\end{figure}

\begin{figure}[h]
\setlength{\unitlength}{5mm}
\centering
\includegraphics[width=3.0in]{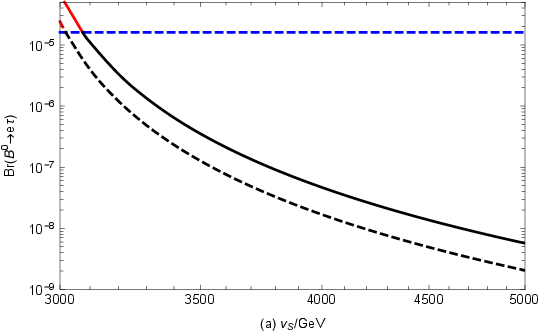}
\vspace{0.2cm}
\setlength{\unitlength}{5mm}
\centering
\includegraphics[width=3.0in]{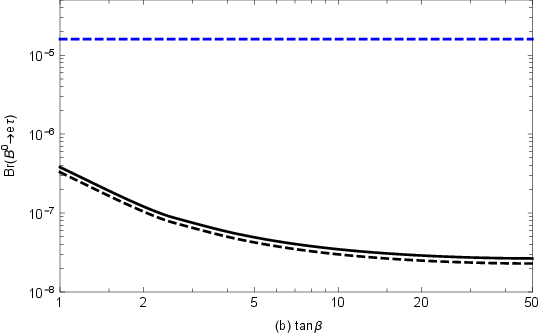}
\caption{ $Br(B^0\rightarrow e{\tau})$  diagrams affected by $v_S$ and $\tan\beta$. The blue dashed line stands for the upper limit on $Br(B^0\rightarrow e{\tau})$ at 90\%CL.. In Fig.\ref {N14}(a), the dashed line (solid line) corresponds to $M^2_{\tilde{L}13} = 300~\rm GeV^2$ ($500~\rm GeV^2$). In Fig.\ref {N14}(b), the dashed line (solid line) corresponds to $M^2_{\tilde{L}13} = 1300~\rm GeV^2$ ($1400~\rm GeV^2$).}\label{N14}
\end{figure}

In this subsection, we analyze the decay with LFV $B^0\rightarrow e{\tau}$ in the $U(1)_X$SSM. To study the influences of different parameters on $Br(B^0\rightarrow e{\tau})$, we suppose the parameters $M^2_{\tilde{L}ii}=1~{\rm TeV}^2$,~$M^2_{{\tilde{\nu}}ii}=0.6~{\rm TeV}^2$,~$M^2_{\tilde{U}nn}=2.6~{\rm TeV}^2$,~(i,n=1,2,3).
In the Fig.\ref {N13} and Fig.\ref {N14}, we picture $Br(B^0\rightarrow e{\tau})$ and $Br({\tau}\rightarrow{e\gamma})$ varying with the different parameters, where
the blue dashed lines denote the latest experimental upper limits of $Br(B^0\rightarrow e{\tau})$ and $Br({\tau}\rightarrow{e\gamma})$.

In the Fig.\ref {N13}(a), we study the branching ratio of $B^0\rightarrow e{\tau}$ versus
$M^2_{\tilde{L}13}$. The line increases with $M^2_{\tilde{L}13}$ increasing from $10^{3}$ $\rm GeV^2$ to $1.5\times10^{4}$ $\rm GeV^2$, which indicates that $M^2_{\tilde{L}13}$ is a sensitive parameter for the numerical results. In the Fig.\ref {N13}(b), we plot $Br({\tau}\rightarrow{e\gamma})$ versus $M^2_{\tilde{L}13}$. As the parameter $M^2_{\tilde{L}13}$ increases, $Br({\tau}\rightarrow{e\gamma})$ will grow rapidly, and $Br({\tau}\rightarrow{e\gamma})$ will exceed the experimental limit. The red solid line indicates the part where the branching ratio of ${\tau}\rightarrow{e\gamma}$ exceeds the upper limit of the experiment. In addition, it is not difficult to see from Fig.\ref {N13}(a) that $Br(B^0\rightarrow e{\tau})$ cannot reach the experimental upper limit under the same parameter space. When we consider the constraint of $Br({\tau}\rightarrow{e\gamma})$ to $Br(B^0\rightarrow e{\tau})$, $Br(B^0\rightarrow e{\tau})$ can be up to $10^{-9}$. In the Fig.\ref {N13}(a), we highlight the part excluded by $Br({\tau}\rightarrow{e\gamma})$ with the green solid line. It can be seen that the limit of $Br({\tau}\rightarrow{e\gamma})$ to $Br(B^0\rightarrow e{\tau})$ is very strict, making it difficult for $Br(B^0\rightarrow e{\tau})$ to reach the upper limit of the experiment.

In the Fig.\ref {N13}(c)-Fig.\ref {N13}(f), we explore the relationship between $m_{\tilde{\nu}_L}$, $\mu$ and the branching ratios of $B^0\rightarrow e{\tau}$, ${\tau}\rightarrow{e\gamma}$ respectively. In these pictures, we set $\tan\beta=10$, $\mu=1200~{\rm GeV}$, $v_S=4.3~\rm TeV$ and $m_{\tilde{\nu}_L}=800~{\rm GeV}$, unless it is a variable in a graph. It is not hard to find that all four figures show a downward trend, that is, $Br(B^0\rightarrow e{\tau})$ and $Br({\tau}\rightarrow{e\gamma})$ decrease with the increase of $m_{\tilde{\nu}_L}$ and $\mu$. The influence of $m_{\tilde{\nu}_L}$ and $\mu$ on $B^0\rightarrow e{\tau}$ is more obvious than that of $Br({\tau}\rightarrow{e\gamma})$, which fails to make the two processes reach the experimental upper limit. In the Fig.\ref {N13}(c), we can clearly see that when the parameters $m_{\tilde{\nu}_L}$ close $1400~{\rm GeV}$ and $M^2_{\tilde{L}13} = 300~\rm GeV^2$, $B^0\rightarrow e{\tau}$ is very close to the experimental upper limit.

In the Fig.\ref {N14}, we explore the effects of $v_S$ and $\tan\beta$ on $Br(B^0\rightarrow e{\tau})$. The effects of $v_S$ and $\tan\beta$ on $Br({\tau}\rightarrow{e\gamma})$ are not large, but the effect on $Br(B^0\rightarrow e{\tau})$ is relatively obvious. We first set
appropriate numerical values for the relevant parameters, such as $\tan\beta=20$, $\mu=1200~{\rm GeV}$, $v_S=3.9~\rm TeV$ and $m_{\tilde{\nu}_L}=800~{\rm GeV}$, unless it is a variable in a graph. The influences of $v_S$ and $\tan\beta$ on $Br(B^0\rightarrow e{\tau})$ show a downward trend, that is, $Br(B^0\rightarrow e{\tau})$ decreases with the increase of $v_S$ and $\tan\beta$. We plot $Br(B^0\rightarrow e{\tau})$ versus $v_S$, in which the dashed line corresponds to $M^2_{\tilde{L}13} = 300~\rm GeV^2$ and the solid line corresponds to $M^2_{\tilde{L}13} = 500~\rm GeV^2$ in Fig.~\ref{N14}(a). In the diagram, we use red solid and red dashed lines to mark the part of $Br(B^0\rightarrow e{\tau})$ that exceeds the experimental limit. In the Fig.\ref {N14}(b), we plot $Br(B^0\rightarrow e{\tau})$ versus $\tan\beta$, in which the dashed line corresponds to $M^2_{\tilde{L}13} = 1300~\rm GeV^2$ and the solid line corresponds to $M^2_{\tilde{L}13} = 1400~\rm GeV^2$. It can be seen that the overall values satisfy the limit. In the Fig.\ref {N14}, with the lines in the figure go from bottom to top, the value of $M^2_{\tilde{L}13}$ increases gradually.

Furthermore, we also investigate the effects of parameters $\Delta$, $g_{YX}$ and $M^2_{\tilde{L}}$ on the branching ratio of process $B^0\rightarrow e{\tau}$. We set $\Delta^{LL}_{nk}(\tilde{U}),~\Delta^{RR}_{nk}(\tilde{U})$ and $\Delta^{LL}_{ij}(\tilde{D})$ ($n\ne k$, $i\ne j$) to be equal and their values are represented by $\Delta$. Through numerical analysis, we find that parameters $\Delta$, $g_{YX}$ and $M^2_{\tilde{L}}$ have a very weak influence on $Br(B^0\rightarrow e{\tau})$, so we will not show the diagram here. The analysis shows that the influence of $\Delta$ on $B^0\rightarrow e{\tau}$ shows an increasing trend. In other words, as $\Delta$ increases, $Br(B^0\rightarrow e{\tau})$ also increases slightly. The influences of $g_{YX}$ and $M^2_{\tilde{L}}$ on $Br(B^0\rightarrow e{\tau})$ show a decreasing trend. In other words, as $g_{YX}$ and $M^2_{\tilde{L}}$ increases, $Br(B^0\rightarrow e{\tau})$ decreases slightly.

From the Fig.\ref {N13} and Fig.\ref {N14}, we can conclude that $M^2_{\tilde{L}13}$, $m_{\tilde{\nu}_L}$, $\mu$, $v_S$ and $\tan\beta$ are sensitive parameters to $Br(B^0\rightarrow e{\tau})$. According to the numerical results of the process, we can also analyze the influence of these parameters on LFV. Similar to our previous analysis, the parameter $M^2_{\tilde{L}13}$ can amplify the effect of LFV, and the parameters $\mu$, $m_{\tilde{\nu}_L}$, $v_S$ and $\tan\beta$ reduce the effect of LFV.
\subsection{$B^0\rightarrow {\mu}{\tau}$}
The experimental upper bound for the LFV process $B^0\rightarrow {\mu}{\tau}$ is $1.2\times10^{-5}$, which is in the same order of magnitude as the process $B^0\rightarrow e{\tau}$. In this part, we still explore the influence of different parameters on the branching ratio of $B^0\rightarrow {\mu}{\tau}$, including the parameters $M^2_{\tilde{L}23}$, $m_{\tilde{\nu}_L}$, $\mu$, $v_S$, $\tan\beta$, $\Delta$, $g_{YX}$ and $M^2_{\tilde{L}}$.

\begin{figure}[h]
\setlength{\unitlength}{5mm}
\centering
\includegraphics[width=3.0in]{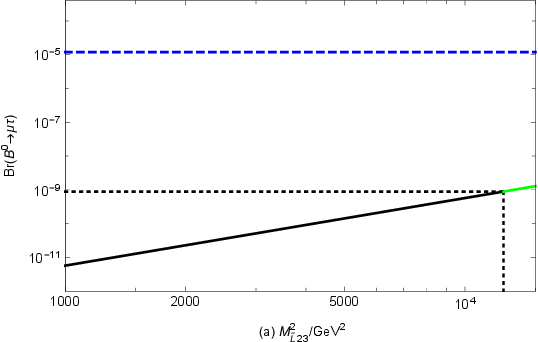}
\vspace{0.2cm}
\setlength{\unitlength}{5mm}
\centering
\includegraphics[width=3.0in]{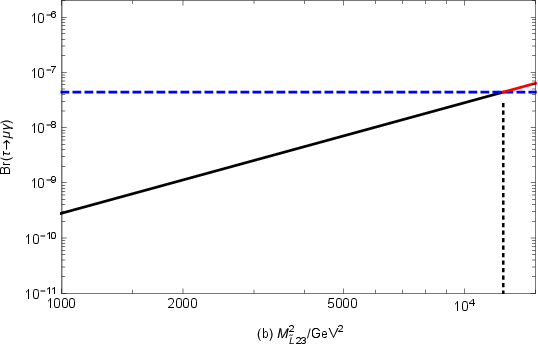}
\vspace{0.2cm}
\setlength{\unitlength}{5mm}
\centering
\includegraphics[width=3.0in]{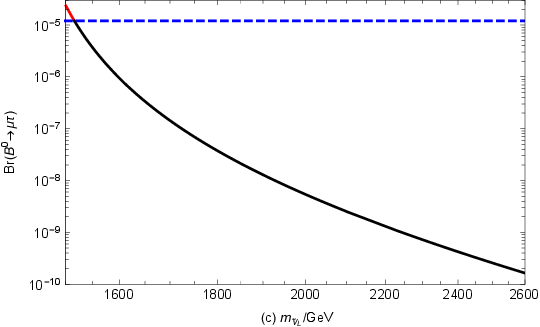}
\vspace{0.2cm}
\setlength{\unitlength}{5mm}
\centering
\includegraphics[width=3.0in]{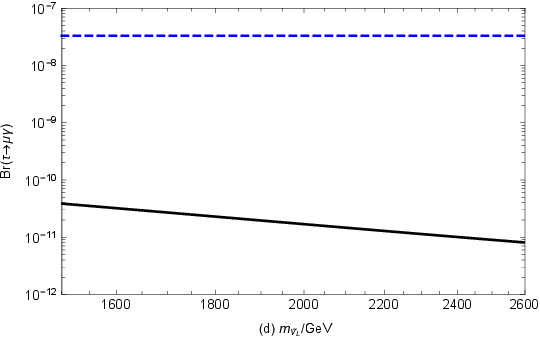}
\vspace{0.2cm}
\setlength{\unitlength}{5mm}
\centering
\includegraphics[width=3.0in]{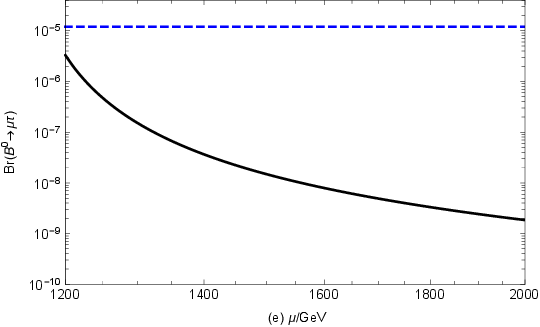}
\vspace{0.2cm}
\setlength{\unitlength}{5mm}
\centering
\includegraphics[width=3.0in]{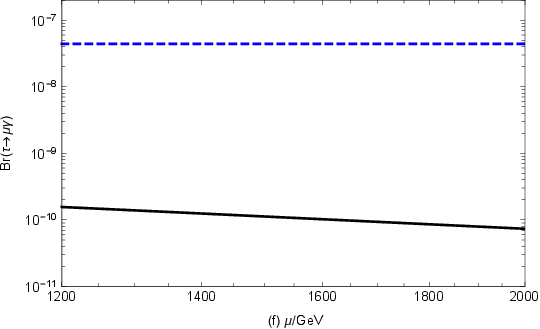}
\caption{ $Br(B^0\rightarrow {\mu}{\tau})$ and $Br({\tau}\rightarrow{{\mu}\gamma})$  diagrams affected by different parameters. The blue dashed line stands for the upper limit on $Br(B^0\rightarrow {\mu}{\tau})$ and $Br({\tau}\rightarrow{{\mu}\gamma})$ at 90\%CL.. }\label{N15}
\end{figure}
We still consider the influence of different parameters on $Br(B^0\rightarrow {\mu}{\tau})$ first, and then
we consider the restriction from the experimental upper limit of rare process $Br({\tau}\rightarrow{{\mu}\gamma})$ on process $Br(B^0\rightarrow {\mu}{\tau})$.
In the Fig.\ref {N15}(a) and Fig.\ref {N15}(b),
we set the parameters $\tan\beta=25$, $\mu=500~{\rm GeV}$, $v_S=3.8~\rm TeV$. We plot the branching ratios of $B^0\rightarrow {\mu}{\tau}$ and ${\tau}\rightarrow{{\mu}\gamma}$ versus $M^2_{\tilde{L}23}$. Fig.\ref {N15}(a) and Fig.\ref {N15}(b) show that the branching ratios of $B^0\rightarrow {\mu}{\tau}$ and ${\tau}\rightarrow{{\mu}\gamma}$ increase with the increase of $M^2_{\tilde{L}23}$. $Br({\tau}\rightarrow{{\mu}\gamma})$ will exceed the experimental limit,  and the red solid line indicates the part where the branching ratio of ${\tau}\rightarrow{{\mu}\gamma}$ exceeds the upper limit of the experiment.
It is not difficult to see that $Br(B^0\rightarrow {\mu}{\tau})$ cannot reach the experimental upper limit under the same parameter space. When we consider the constraint of $Br({\tau}\rightarrow{{\mu}\gamma})$ to $Br(B^0\rightarrow {\mu}{\tau})$, $Br(B^0\rightarrow {\mu}{\tau})$ can be up to $10^{-9}$. In the Fig.\ref {N15}(a), we highlight the part excluded by $Br({\tau}\rightarrow{{\mu}\gamma})$ with the green solid line. It can be seen that the limit of $Br({\tau}\rightarrow{{\mu}\gamma})$ to $Br(B^0\rightarrow {\mu}{\tau})$ is very strict, making it difficult for $Br(B^0\rightarrow {\mu}{\tau})$ to reach the upper limit of the experiment.

In the Fig.\ref {N15}(c)-Fig.\ref {N15}(f), we explore the relationship between
$m_{\tilde{\nu}_L}$, $\mu$ and the branching ratios of
$B^0\rightarrow {\mu}{\tau}$, ${\tau}\rightarrow{{\mu}\gamma}$ respectively,
with the parameters $\tan\beta=20$, $\mu=1200~{\rm GeV}$, $v_S=3.9~\rm TeV$
and $m_{\tilde{\nu}_L}=800~{\rm GeV}$.
It is not hard to find that all four figures show a downward trend, that is, $Br(B^0\rightarrow {\mu}{\tau})$ and $Br({\tau}\rightarrow{{\mu}\gamma})$ decrease with the increases of $m_{\tilde{\nu}_L}$ and $\mu$. The influences of $m_{\tilde{\nu}_L}$ and $\mu$ on $B^0\rightarrow {\mu}{\tau}$ are more obvious than that of $Br({\tau}\rightarrow{{\mu}\gamma})$. In the Fig.\ref {N15}(c), $Br(B^0\rightarrow {\mu}{\tau})$ can still reach the experimental upper limit, but $Br(B^0\rightarrow {\mu}{\tau})$ and $Br({\tau}\rightarrow{{\mu}\gamma})$ cannot do that in the Fig.\ref {N15}(d)-Fig.\ref {N15}(f).

\begin{figure}[h]
\setlength{\unitlength}{5mm}
\centering
\includegraphics[width=3.0in]{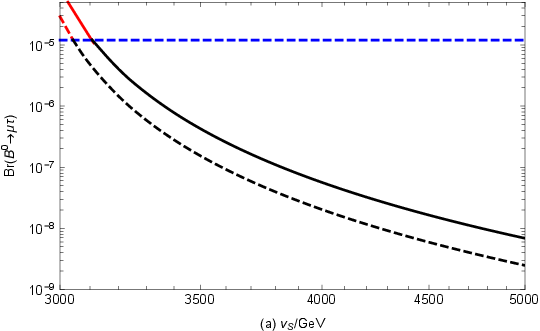}
\vspace{0.2cm}
\setlength{\unitlength}{5mm}
\centering
\includegraphics[width=3.0in]{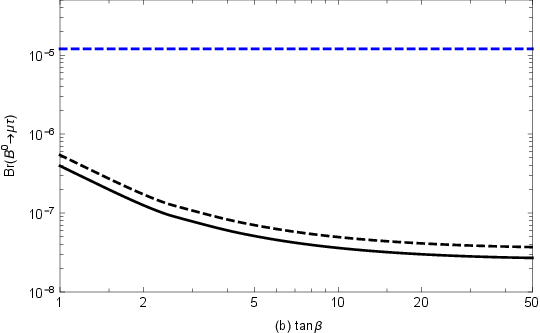}
\caption{ $Br(B^0\rightarrow {\mu}{\tau})$  diagrams affected by $v_S$ and $\tan\beta$. The blue dashed line stands for the upper limit on $Br(B^0\rightarrow {\mu}{\tau})$ at 90\%CL.. In Fig.\ref {N16}(a), the dashed line (solid line) correspond to $M^2_{\tilde{L}23} = 300~\rm GeV^2$ ($500~\rm GeV^2$). In Fig.\ref {N16}(b), the dashed line (solid line) correspond to $v_S = 3.9~\rm TeV$ ($4.0~\rm TeV$).}\label{N16}
\end{figure}

In the Fig.\ref {N16}, we explore the effects of $v_S$ and $\tan\beta$ on $Br(B^0\rightarrow {\mu}{\tau})$. The effect of $v_S$ and $\tan\beta$ on $Br({\tau}\rightarrow{{\mu}\gamma})$ is weak, but the effect on $Br(B^0\rightarrow {\mu}{\tau})$ is relatively obvious. The influences of $v_S$ and $\tan\beta$ on $Br(B^0\rightarrow {\mu}{\tau})$ show a downward trend, that is, $Br(B^0\rightarrow {\mu}{\tau})$ decreases with the increases of $v_S$ and $\tan\beta$. We plot $Br(B^0\rightarrow {\mu}{\tau})$ versus $v_S$, in which the dashed line corresponds to $M^2_{\tilde{L}23} = 300~\rm GeV^2$ and the solid line corresponds to $M^2_{\tilde{L}23} = 500~\rm GeV^2$ in Fig.~\ref{N16}(a). In the diagram, we use red solid and dotted lines to mark the part of $Br(B^0\rightarrow {\mu}{\tau})$ that exceeds the experimental limit. As the lines in the figure go from bottom to top, the value of $M^2_{\tilde{L}23}$ increases gradually. In the Fig.\ref {N16}(b), we plot $Br(B^0\rightarrow {\mu}{\tau})$ versus $\tan\beta$, in which the dashed line corresponds to $v_S = 3.9~\rm TeV$ and the solid line corresponds to $v_S = 4.0~\rm TeV$. It can be seen that the overall values satisfy the limit. As the lines in the figure go from bottom to top, the value of $v_S$ decreases gradually.

Moreover, through numerical analysis, we find that parameters $\Delta$, $g_{YX}$ and $M^2_{\tilde{L}}$ have the weak influence on the branching ratio of process $B^0\rightarrow {\mu}{\tau}$. We set $\Delta^{LL}_{nk}(\tilde{U}),~\Delta^{RR}_{nk}(\tilde{U})$ and $\Delta^{LL}_{ij}(\tilde{D})$ ($n\ne k$, $i\ne j$) to be equal and their values are represented by $\Delta$. We will not show the diagram here. The analysis implies that the influence of $\Delta$ on $B^0\rightarrow {\mu}{\tau}$ shows an increasing trend. In other words, as $\Delta$ increases, $B^0\rightarrow {\mu}{\tau}$ also increases slightly. The influences of $g_{YX}$ and $M^2_{\tilde{L}}$ on $B^0\rightarrow {\mu}{\tau}$ show a decreasing trend. In other words, as $g_{YX}$ and $M^2_{\tilde{L}}$ increases, $B^0\rightarrow {\mu}{\tau}$ decreases slightly.

\subsection{$B^0\rightarrow{M^0{l_i}^{\pm}{l_j}^{\mp}}$, $B^+\rightarrow{M^+{l_i}^{\pm}{l_j}^{\mp}}$, with $M = \pi, \rho$}
In this paper, we use the general expressions provided in \cite{ff4,FB1} to perform a brief numerical analysis of the corresponding process using our method. We set different parameters as researched variables, including $\tan\beta$, $\mu$, $m_{\tilde{\nu}_L}$, $v_S$.
When $\tan\beta$=25, $\mu$=1200 GeV, $m_{\tilde{\nu}_L}$=800 GeV and $v_S$=3.8 TeV,
the branching ratios of processes $B^0\rightarrow {\rho^0}e{\mu}$ and $B^+\rightarrow {\pi^+}e{\mu}$ can reach $10^{-8}$ and the branching ratios of processes $B^0\rightarrow {\rho^0}l{\tau}$ and $B^+\rightarrow {\pi^+}l{\tau}$ ($l = e,{\mu}$) can reach $10^{-9}$.

\section{discussion and conclusion}

 In summary, we have investigated  the LFV process $B^0\rightarrow{{l_i}^{\pm}{l_j}^{\mp}}$ ($B^0\rightarrow e{\mu}$, $B^0\rightarrow e{\tau}$, and $B^0\rightarrow {\mu}{\tau}$) with the method of MIA in the $U(1)_X$SSM.
Thinking about the box-type diagrams associated with this process, we get the numerical values of degenerated results and draw two-dimensional plots from the large number of numerical results obtained. During our analysis of numerical results, we set many different parameters as researched variables, including $\tan\beta$, $g_{YX}$, $\mu$, $v_S$, $m_{\tilde{\nu}_L}$, $M^2_{\tilde{L}}$, $M^2_{\tilde{L}ij}$ $\delta_{ij}^{LL}$ and $\Delta$ $(i\ne j)$. By analyzing the numerical results of the adopted parameter space, we can conclude that $\mu$, $v_S$, $m_{\tilde{\nu}_L}$, $M^2_{\tilde{L}ij}$ $(i\ne j)$ are the sensitive parameters with great influence on the branching ratio of the process $B^0\rightarrow{{l_i}^{\pm}{l_j}^{\mp}}$.

The off-diagonal elements $M^2_{\tilde{L}ij}~(i\neq j, ~i,~j=1,~2,~3)$ emerge in the mass squared matrixes of slepton and sneutrino,
  where $i$ and $j$ correspond to generations of the final leptons $l_i$ and $l_j$.
    $M^2_{\tilde{L}ij}$ are still the main sensitive parameters and LFV sources.  From the analysis of our numerical results, we find that the branching ratio of $B^0\rightarrow e{\mu}$ can reach $10^{-12}$. The branching ratios of $B^0\rightarrow e{\tau}$ and $B^0\rightarrow {\mu}{\tau}$ can reach $10^{-7}-10^{-8}$. Most of the explored parameters can satisfy the experimental limit or break the upper limit of the experiment, which provide new ideas for finding NP.

    In our previous work, we have researched the LFV decays $l_j\rightarrow{l_i\gamma}$ with the method of MIA in the $U(1)_X$SSM. For processes $B^0\rightarrow{{l_i}^{\pm}{l_j}^{\mp}}$ ($B^0\rightarrow e{\mu}$, $B^0\rightarrow e{\tau}$ and $B^0\rightarrow {\mu}{\tau}$) and $l_j\rightarrow{l_i\gamma}$ (${\mu}\rightarrow{e\gamma}$, ${\tau}\rightarrow{e\gamma}$ and ${\tau}\rightarrow{{\mu}\gamma}$), we have investigated and found that as parameters $M^2_{\tilde{L}12}$, $M^2_{\tilde{L}13}$ and $M^2_{\tilde{L}23}$ increase, $Br(B^0\rightarrow{{l_i}^{\pm}{l_j}^{\mp}})$ and $Br(l_j\rightarrow{l_i\gamma})$ increase rapidly. $M^2_{\tilde{L}ij}$ are
 LFV sources for the decays $B^0\rightarrow{{l_i}^{\pm}{l_j}^{\mp}}$ and $l_j\rightarrow{l_i\gamma}$.
Also as parameter $m_{\tilde{\nu}_L}$ increases, $Br(B^0\rightarrow{{l_i}^{\pm}{l_j}^{\mp}})$ and $Br(l_j\rightarrow{l_i\gamma})$ rapidly decrease.
The reason should be that heavy particle mass suppresses the NP contribution to LFV processes. From the comparison of these processes we can see that the parameters $M^2_{\tilde{L}12}$, $M^2_{\tilde{L}13}$, $M^2_{\tilde{L}23}$ and $m_{\tilde{\nu}_L}$ have a strong influence on $Br(B^0\rightarrow{{l_i}^{\pm}{l_j}^{\mp}})$ and $Br(l_j\rightarrow{l_i\gamma})$. Considering the constraint from $Br(l_j\rightarrow{l_i\gamma})$ to $Br(B^0\rightarrow{{l_i}^{\pm}{l_j}^{\mp}})$,
we explore and find that the limit from $Br(l_j\rightarrow{l_i\gamma})$ to $Br(B^0\rightarrow{{l_i}^{\pm}{l_j}^{\mp}})$ is very strict. So, it is difficult for $Br(B^0\rightarrow{{l_i}^{\pm}{l_j}^{\mp}})$ to reach the experimental upper limits. Perhaps the accuracy of the experiment will be further improved in the near future.

Based on our studies of the relevant model parameters, with the desired future experimental sensitivity for LHCb the LFV decays may be discovered in the future. Through our analysis, flavor mixing parameters ($M^2_{\tilde{L}ij}$ $(i\ne j)$) are very important sensitive parameters that may be effectively tested at future colliders. Perhaps the sensitivity of future experiments will be more significantly improved.

{\bf Acknowledgments}

This work is supported by National Natural Science Foundation of China (NNSFC)
(No. 12075074), Natural Science Foundation of Hebei Province
(A2023201041, A202201022, A2022201017, A2023201040), Natural Science Foundation of Hebei Education Department (QN2022173), Post-graduate's Innovation Fund Project of Hebei University (HBU2023SS043), the youth top-notch talent support program of the Hebei Province.

\end{document}